%% file: MCMCNutshellArxiv.tex
\documentclass[prb,reprint, floatfix, 10pt]{revtex4-2}
\usepackage[labelfont=bf,font=small, justification=raggedright]{caption}
\usepackage{pifont}
\usepackage{cancel}
\usepackage{slashbox}
\usepackage{hyperref}
\usepackage{graphicx}
\usepackage[utf8]{inputenc}
\usepackage{amsmath}
\usepackage[normalem]{ulem}
\usepackage{verbatim}
\usepackage{amssymb}
\usepackage{listings}
\usepackage{hyperref}
\usepackage{mathtools}
\usepackage{color}
\usepackage{xspace}
\usepackage{lipsum}
\usepackage{tikz-cd}
\usepackage{float}
\usepackage[all]{nowidow}
\newcommand{\prog}[1]{Alg.~\ref{alg:#1} (\sub{#1})}
\newcommand{\progtest}[1]{\ref{alg:#1} \sub{#1}}

\newcommand{\progg}[1]{Algorithm~\ref{alg:#1} (\sub{#1})}

\newcommand{\progn}[1]{Alg.~\ref{alg:#1}}

\newcommand{\Period}{\tau}
\newcommand{\pit}[1]{\pi^{\{#1\}}} 
\newcommand{\piiso}{\pi^{\text{iso}}} 
\newcommand{\pMet}{\PCAL^{\text{Met} }}
\newcommand{\pFact}{\PCAL^{\text{fact}}}

\newcommand{\Plift}{P^{\text{lift}}}
\newcommand{\xmax}{x_{\max}}
\newcommand{\tevent}{t_{\text{ev}}}
\newcommand{\xeventtwofour}{x_{\text{ev}}}
\newcommand{\xeventtwo}{x_{\text{ev}}^{(2)}}
\newcommand{\xeventfour}{x_{\text{ev}}^{(4)}}
\newcommand{\xeventtwofourhat}{\xhat_{\text{ev}}}
\newcommand{\xeventtwohat}{\xhat_{\text{ev}}^{(2)}}
\newcommand{\xeventfourhat}{\xhat_{\text{ev}}^{(4)}}
\newcommand{\qtwofourhat}{\qhat}
\newcommand{\qtwohat}{\qhat^{(2)}}
\newcommand{\qfourhat}{\qhat^{(4)}}
\newcommand{\MCMCNutshell}{\texttt{MCMCNutshell}\xspace}

\input{latexcommands_arxiv.tex}

\newcommand{\cmark}{\ding{51}}

\newcommand{\Utwo}{U_{2}}
\newcommand{\Ufour}{U_{4}}
\newcommand{\Utwofour}{U_{24}}
\newcommand{\Ubound}{\widehat{U}}
\newcommand{\Uboundtwofour}{\widehat{U}_{24}}
\newcommand{\Uboundtwo}{\widehat{U}_{2}}
\newcommand{\Uboundfour}{\widehat{U}_{4}}
\newcommand{\pitwofour}{\pi_{24}}

\newcommand{\pitwo}{\pi_{2}}

\newcommand{\ZZ}{\mathbb{Z}}

\newcommand{\pifour}{\pi_{4}}

\newcommand\subfig[2]{{Fig.~\ref{#1}{#2}}}
\newcommand\subcap[1]{{(#1):}}

\hypersetup{
    colorlinks=true,
    linkcolor=blue,
    filecolor=magenta,      
    urlcolor=cyan,
    pdftitle={Concepts in Monte Carlo sampling}
}

\begin{document}
\newfloat{algorithm}{ht}{loa}
\floatname{algorithm}{Algorithm }
\setcounter{algorithm}{-1}
\title{Concepts in Monte Carlo sampling}
\author{Gabriele Tartero}
\author{Werner Krauth}
\email{werner.krauth@ens.fr}
\date{\today}
\affiliation{Laboratoire de Physique de l’Ecole normale supérieure, ENS,
Université PSL, CNRS, Sorbonne Université, Université Paris-Diderot, Sorbonne
Paris Cité, Paris, France}
\begin{abstract}
We discuss modern ideas in Monte Carlo algorithms in the simplified setting of
the one-dimensional anharmonic oscillator.
After reviewing the connection between molecular dynamics and
Monte Carlo, we introduce to the Metropolis and the factorized
Metropolis algorithms and to lifted non-reversible Markov chains. We
furthermore illustrate the concept of thinning,
where moves are accepted by
simple bounding potentials rather than, in our case, the harmonic and quartic
constituents of the anharmonic oscillator. We point out the multiple
connections of our example algorithms with real-world sampling problems.
The paper is fully self-contained and Python implementations are provided.
\end{abstract}
\maketitle
    
\section{Introduction}
The Monte Carlo
method is an important tool for producing samples $x$ from a given probability
distribution $\pi(x)$. In real-life applications, algorithms and computer
implementations for this sampling
problem can be highly complex. In this paper, we rather discuss a dozen of
distinct
Monte Carlo algorithms in the severely stripped-down setting of a particle in a
one-dimensional anharmonic potential
\begin{equation}
    \Utwofour(x) = \fracb{x^2}{2} + \frac{x^4}{4}
\label{equ:potential}
\end{equation}
consisting of a  harmonic term, $\Utwo = x^2/2$, and
a quartic one, $\Ufour = x^4/4$. For concreteness, we also provide short
example programs.
\begin{figure}[htb]
    \centering
    \includegraphics{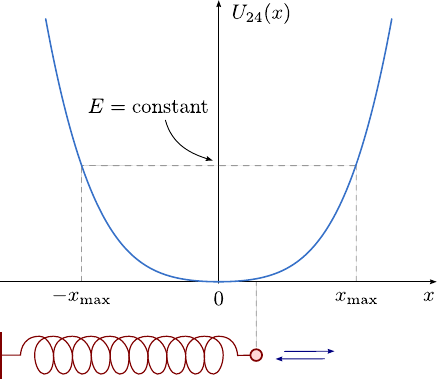}
    \caption{Isolated anharmonic oscillator at energy $E$,
    subject to the potential $\Utwofour$ of \Eq{equ:potential}.}
\label{fig:IsolatedOscillator}
\end{figure}

For the anharmonic oscillator, the distribution to be sampled is the
Boltzmann distribution
\begin{equation}
    \pitwofour(x) = \expc{-\beta \Utwofour(x)},
\label{equ:BoltzmannDistribution}
\end{equation}
where $ \beta = (k_B T)^{-1} $ is the inverse of the temperature $T$, and
$k_B $ denotes the Boltzmann constant.
The connection between the potential $\Utwofour$ and the distribution
$\pitwofour$ derives from the following. In classical mechanics, an isolated
particle is governed by Newton's law and, in a one-dimensional confining
potential, oscillates between two turning points. A certain function $\piiso(x)$
describes the fraction of time that the particle spends at position $x$ during
one period and, therefore, during a long time interval containing many periods.
If the particle is in contact with a thermostat, this function turns into a
probability distribution for finding the particle at a position $x$ at large
times $t$, and it is exactly the Boltzmann distribution $\pitwofour(x)$ of
\Eq{equ:BoltzmannDistribution}, as we will discuss (see
\sect{sec:ClassicalToStatistical}). The molecular-dynamics method generally accesses this
distribution through the numerical solution of Newton's equation in contact 
with a thermostat.

The Monte Carlo method addresses the sampling problem more abstractly than
molecular dynamics, as it samples (obtains samples $x$ from) the
distribution $\pitwofour(x)$ without simulating a physical process. The
sequence of twelve short yet intricate Monte Carlo algorithms that we present
here will lead us from the beginning of the method, namely direct sampling and
the reversible Metropolis algorithm and its extensions
(\sect{sec:reversible_markov_chains}), to non-reversible Markov-chain algorithms
(\sect{sec:NonReversible}) and to advanced approaches that sample the target
distribution with a minimum of evaluations of the potential
(\sect{sec:Thinning}). Some mathematical results are collected separately
(\app{app:MathDetails}). Our algorithms are presented in compact pseudo-code
(as in~\cite{SMAC}) and implemented in short, openly accessible, Python
programs (\app{app:Software}). Their correctness is tested 
to high precision (\app{app:Tests}). A companion 
paper~\cite{Tartero2023b} will translate the concepts discussed here to
real-life settings and address efficiency questions whereas in the present
paper, we are only concerned with the correctness of the sampling algorithms.

\section{From classical to statistical mechanics}
\label{sec:ClassicalToStatistical}

If isolated from the environment, so that the energy is conserved, the
anharmonic
oscillator of \fig{fig:IsolatedOscillator} is a classical, periodic,
one-dimensional deterministic system, and we may track the fraction of time per
period that the particle spends near a given position $x$
(\subsect{subsec:MolecularDynamicsE}). When interacting with a heatbath (for
which we suppose a concrete realization), the motion is piecewise
deterministic~\cite{Davis1984},
yet random. In this case, we may sample the Boltzmann distribution 
$\pitwofour$ through
a molecular-dynamics modeling of the particle subject to Newton's laws and 
interacting with the thermostat (\subsect{subsec:MolecularDynamicsT}). At
the end of this section, we provide a Monte Carlo algorithm that directly
samples $x$ from the Boltzmann distribution (\subsect{subsec:DirectSampling}).
 
\subsection{The isolated anharmonic oscillator}
\label{subsec:MolecularDynamicsE}
\begin{figure}[htb]
    \centering
    \includegraphics{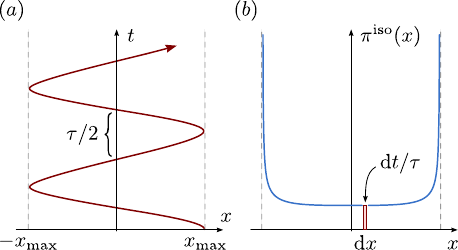}
    \caption{Isolated anharmonic oscillator, as represented in
\fig{fig:IsolatedOscillator}. \subcap{a} Periodic trajectory with amplitude
$2x_{\max}$ and period $\Period$. \subcap{b} Normalized function $\piiso$. The
fraction of time $\diff t /\Period $ spent per period between $x$ and $x + \diff
x$ is $\piiso(x) \diff x$.}
\label{fig:Landau-Lifshitz}
\end{figure}

We may hold the  particle fixed---with velocity $v=0$---then
release it at time $t=0$ from a position $x_{\max} > 0$. If it is isolated,
the anharmonic oscillator conserves its energy $E$, given by the sum
of the kinetic and  potential energies at all times $t \ge 0$. It thus picks up 
velocity until it reaches
the minimum of the potential at $x=0$, then slows down and turns around at
$-x_{\max}$, where $E$ equals the potential energy and
the velocity again vanishes (see \subfig{fig:Landau-Lifshitz}{a}). The energy 
$E$ is then
\begin{equation}
E = \fracb{ \xmax^2}{2} + \fracb{\xmax^4}{4} \Leftrightarrow
x_{\max} = \sqrt{-1 + \sqrt{1 + 4E}},
\label{equ:amplitude}
\end{equation}
as follows from solving a quadratic equation and taking a square root.
In between the turning points $-\xmax$ and $\xmax$ the kinetic energy 
$\half (\diff x / \diff t)^2$ is positive, and the conservation of energy can be 
written as
\begin{equation}
    E = \fracb{1}{2} \glb \fracb{\diff x}{\diff t} \grb ^ 2\!\!\! + \Utwofour(x)
\Leftrightarrow
\fracb{\diff x}{\diff t} = \pm \sqrt{2\glc E - \Utwofour(x) \grc},
\label{equ:energy}
\end{equation}
which gives
\begin{equation}
    \diff t = \pm \sqrt{\fracb{1}{2 \glc E - \Utwofour(x) \grc }}  \diff x.
\label{equ:RatioIso}
\end{equation}

\begin{figure}[htb]
    \centering
    \includegraphics{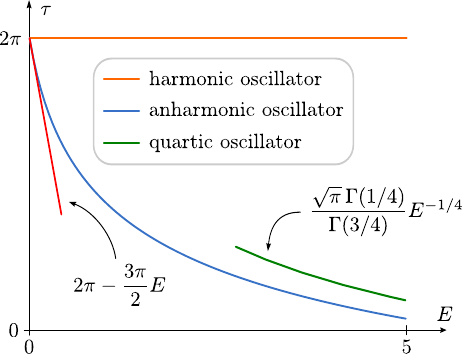}
\caption{Period $\Period$ of the isolated anharmonic oscillator as a function
of the energy $E$. The period of the harmonic oscillator is independent of $E$, 
while that of the quartic oscillator scales as $E^{-1/4}$. Here, $\Gamma$ 
denotes the Euler gamma function (see \app{app:MathDetails}).}
\label{fig:period_vs_energy}
\end{figure}

The period $\tau$ of the motion, $\ie$ the time between two realizations
of a given position and velocity, corresponds to four times the interval from 
$x=0$ to $\xmax$,
\begin{multline}
 \Period = 4 \int_0^{\xmax} \!\!\!\! \diff t  = 4 \int_0^{ \sqrt{-1 + \sqrt{1 +
4E}}}
\fracb{1}{\sqrt{2 \glc E -
\Utwofour(x) \grc}} \diff x \\
         = 4 \sqrt{\fracb{2}{1 + \sqrt{1 + 4 E}}} \, K \glb
        \fracb{1 - \sqrt{1 + 4 E}}{1 + \sqrt{1 + 4 E}}
        \grb,
\label{equ:landau_distr}
\end{multline}
where $K$ is the complete elliptic integral of the first kind (see
\fig{fig:period_vs_energy}).
For small $E$, the period $\Period$ agrees with that of the harmonic
oscillator, which is famously independent of $\xmax$, thus of $E$. For
large $E$, in contrast, the period $\Period \sim  E^{-1/4}$
approaches that of the quartic oscillator (see \app{app:MathDetails} for some 
mathematical details).

\Eqq{equ:RatioIso} yields the fraction $\piiso(x)$ of time that the particle 
spends between $x$ and $x+ \diff x$ over a semi-period,
\begin{equation}
 \piiso(x) = \fracb{2}{\Period} \sqrt{\fracb{1}{2 \glc E - \Utwofour(x) \grc
}},
\end{equation}
with $-\xmax < x < 
\xmax$. The function $\piiso(x)$ is normalized, but it does not represent 
the probability for the particle to be at $x$ at a fixed time $t$, 
because of the deterministic nature of the motion (see 
\subfig{fig:Landau-Lifshitz}{b}).

To simulate the isolated anharmonic oscillator, we could numerically
integrate the first-order ordinary differential equation on the right
of \Eq{equ:energy} 
over a quarter period and then piece together the entire trajectory of
\subfig{fig:Landau-Lifshitz}{a}. However, this method is specific to  
one-dimensional
dynamical systems~\cite[\S 11]{LandauMechanics1982}. 
In order to reflect the general case, 
we numerically integrate Newton's law for the force $F$:
\begin{equation}
F = m \fracd{\diff^2}{\diff x^2} x(t), \quad \text{with } F = -
\fracd{\diff \Utwofour}{\diff x} = -x - x^3.
\end{equation}
Substituting the time differential $\diff t$  by a very small finite interval
$\Delta t$, appropriate for stepping through, from $t$ to $t+ \Delta
t$, and to $t + 2\Delta t$, and so on, we obtain
\begin{align}
x(t + \Delta t) &= x(t) + v(t) \Delta t,\\
v(t + \Delta t) &= v(t) - (x + x^3) \Delta t.
\end{align}
\prog{isolated-dynamics}
implements one iteration of this naive algorithm, that we set off with an
initial
position $x(t=0) = \xmax$, and an initial velocity $v(t=0) = 0$. The output can 
then be fed back
into the input of the program.
As most isolated-molecular-dynamics codes,
\progn{isolated-dynamics} is unstable---the energy will slowly increase with
time, then diverge. To obtain good approximate results, we should use a small 
discretization $\Delta t$ and not
run the program up to excessively large values of $t$.

\begin{algorithm}
    \newcommand{\algo}{isolated-dynamics}
    \captionsetup{margin=0pt,justification=raggedright}
    \begin{center}
        $\begin{array}{ll}
            & \PROCEDURE{\algo}\\
            & \INPUT{x, v, t}\\
            & \IS{t}{t + \Delta t}\\
            & \IS{x'}{x + v \Delta t}\\
            & \IS{v}{v - \glb x + x^3 \grb \Delta t}\\
            & \IS{x}{x'}\\
            & \OUTPUT{x, v, t}\\
            & \ENDPROCEDURE\
        \end{array}$
    \end{center}
    \caption{\sub{\algo}.
    Naive integration of Newton's equations for the
    isolated anharmonic oscillator.
    }
\label{alg:\algo}
\end{algorithm}

\subsection{Introducing a thermal bath}
\label{subsec:MolecularDynamicsT}

\begin{figure}[htb]
    \centering
    \includegraphics{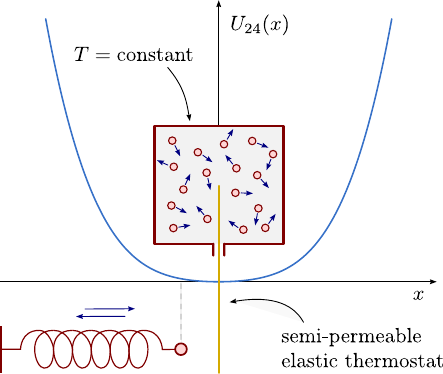}
    \caption{Anharmonic oscillator of \Eq{equ:potential} interacting with a
heatbath at temperature $T$ through an elastic semi-permeable thermostat
vibrating in an infinitesimal interval about $x=0$.}
\label{fig:Heat-bathOscillator}
\end{figure}

Liquids, gases and other systems described by statistical mechanics are
generally composed of particles that interact and exchange energy and momentum.
Any sub-system interacts with its environment and therefore
does not conserve energy
and momentum. For the anharmonic oscillator, this may be modeled by an external
heatbath at temperature $T$, represented by a box composed of a very large number of 
hard-sphere particles of mass
$m=1$ that fly about randomly with velocities given by the Maxwell 
distribution. For concreteness, we imagine the
anharmonic oscillator to be in contact with the heatbath through  a
semi-permeable elastic \quot{thermostat}, a stick that vibrates back and forth 
in an
infinitesimal interval around $x=0$, and that is also of mass one. At each
collision of the thermostat with a heatbath particle, their two velocities
are exchanged. We may imagine that the anharmonic oscillator, as it approaches
$x=0$, passes through the thermostat without interaction with
probability $1/2$, and otherwise bounces off with the velocity of the stick.
The particle trajectory is then deterministic except at the origin (see 
\fig{fig:Heat-bathOscillator}). Statistical
mechanics teaches us that, although all the particles in the heatbath are
Maxwell-distributed, the thermostat behaves differently.
In particular, since the latter lies at a fixed position (up to an 
infinitesimal interval), its velocity follows the distribution
\begin{equation}
    \pi(v) \ddd{v} = \beta |v| \expa{-\fraca{\beta
    v^2}{2}}  \ddd{v},
\label{equ:maxwell_boundary_conditions}
\end{equation}
often called the Maxwell boundary condition (see~\cite[Sec. 2.3.1]{SMAC}).
It differs by the prefactor $\beta |v|$ from the Maxwell distribution of one
velocity component.

\begin{figure}[htb]
    \centering
    \includegraphics{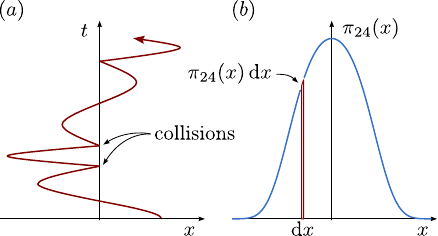}
    \caption{Anharmonic oscillator in contact with the thermostat of
    \fig{fig:Heat-bathOscillator}. \subcap{a} Piecewise deterministic trajectory
with random kicks at $x=0$. \subcap{b} At large $t$, when the initial
configuration $x(t=0)$ is forgotten, the particle position follows 
the Boltzmann distribution $\pitwofour$ of \Eq{equ:BoltzmannDistribution}. }
\label{fig:ThermalOscillator}
\end{figure}

The velocity distribution of the thermostat in
\Eq{equ:maxwell_boundary_conditions}
can be sampled as
\begin{equation}
    v = \pm \sqrt{\fracd{-2 \loga{\ranb{0, 1}}}{\beta}},
\label{equ:membrane_velocity}
\end{equation}
and the Maxwell boundary condition, thus realized with a single random number,
exactly represents the infinite heatbath of \fig{fig:Heat-bathOscillator}.
After a few collisions (see \subfig{fig:ThermalOscillator}{a}),
the particle has forgotten its initial position $x(0)$, and it makes sense to
speak of the probability distribution at time $t$.
Exactly given by $\pitwofour(x)$ in the limit $\Delta t \to 0$, it
substantially differs  from $\piiso$ of
\subfig{fig:Landau-Lifshitz}{b} and is naively sampled by 
\prog{thermostat-dynamics}.

\begin{algorithm}
    \newcommand{\algo}{thermostat-dynamics}
    \captionsetup{margin=0pt,justification=raggedright}
    \begin{center}
        $\begin{array}{ll}
            & \PROCEDURE{\algo}\\
            & \INPUT{x, v, t}\\
            & \IS{x'}{x + v \Delta t}\\
            & \IS{t}{t + \Delta t}\\
            & \IS{\Upsilon}{\ranb{0, 1}}\\
            & \IF{xx' < 0\ \AND\ \Upsilon < \fraca{1}{2}}\\
            & \BRACE{
                     \IS{v}{-\sub{sign}(v) \sqrt{-2 \beta^{-1} \loga{\ranb{0,
                     1}}}} \quad \COMMENT{see
\Eq{equ:membrane_velocity}}
                    }\\
            & \ELSE\\
            & \BRACE{
            \IS{v}{ v - (x + x ^3) \Delta t}\\
            \IS{x}{x'}
            }\\
            & \OUTPUT{x, v, t}\\
            & \ENDPROCEDURE\
        \end{array}$
    \end{center}
    \caption{\sub{\algo}.
    Naive solution of Newton's equations for the anharmonic oscillator with the
semi-permeable thermostat at $x=0$ (see \fig{fig:Heat-bathOscillator}).}
\label{alg:\algo}
\end{algorithm}

We pause for a moment to compute the normalization $Z(\beta)$ of $\pitwofour$ in
\Eq{equ:BoltzmannDistribution}, that is, the partition function
\begin{equation}
    Z(\beta) = \int_{-\infty}^{\infty }
    \dd{x} \pitwofour(x)
    = \fracb{\expa{\beta/8}}{\sqrt{2}}
    K_{1/4} (\beta/8), 
\label{equ:PartitionAnharmonic}
\end{equation}
where $K_{1/4}$ denotes the Bessel function of the second kind (see 
\app{app:MathDetails}). For simplicity of notation, the division by the
partition function is understood whenever we want $\pitwofour$ to represent
a \emph{bona fide} normalized probability distribution.

\subsection{Direct Monte Carlo sampling}
\label{subsec:DirectSampling}
To sample the distribution $\pitwofour$, one need not simulate a physical
system---in our case the anharmonic oscillator in contact with a heatbath.
Let us first consider the simpler problem of the Gaussian distribution:
\begin{equation}
\pitwo(x) = \expc{-\beta \Utwo(x)} =  \expb{- \beta x^2/2}.
\label{equ:GaussianBell}
\end{equation}
Samples $x$ of $\pitwo(x)$ are known as Gaussian random numbers of zero mean
and of standard deviation $1/\sqrt{\beta}$.
They are readily available on computers, websites, and even pocket
calculators (see~\cite[Sec. 1.2.5]{SMAC} for an algorithm using the
method of sample transformation from uniform random numbers).
With an additional uniform random number
$y=\ranc{0, \expb{-\beta x^2/2}}$, they can be expanded into two-dimensional
positions $(x, y)$ of \quot{pebbles} which are uniformly distributed in the
area between the $x$-axis and the bell-shaped Gaussian curve of
\Eq{equ:GaussianBell}.

\begin{figure}[htb]
\centering
\includegraphics{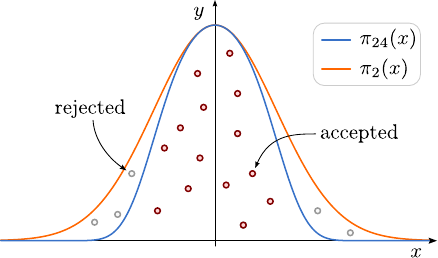}
\caption{Uniformly distributed pebbles below the Gauss curve $\pitwo$. The
$x$-values of the pebbles $(x,y)$, with $y < \pitwofour(x)$, sample the
Boltzmann distribution $\pitwofour$ of \Eq{equ:BoltzmannDistribution}.
}
\label{fig:pi2_vs_pi24}
\end{figure}
The distribution of the anharmonic oscillator satisfies $\pitwofour(x) \le
\pitwo(x)$ for all $x$
(see \fig{fig:pi2_vs_pi24}), and it plays no role that it may not be normalized.
Those pebbles that lie below $\pitwofour$---as
they are uniformly distributed below the Gauss curve---are also evenly spread
out below $\pitwofour$.
Clearly, it suffices to reject any pebble $(x,y)$ above
$\pitwofour$,
to be left with  $x$ positions distributed according to the Boltzmann
distribution of the anharmonic oscillator. \progg{direct-sampling} implements
this direct-sampling idea.
\begin{algorithm}
    \newcommand{\algo}{direct-sampling}
    \captionsetup{margin=0pt,justification=raggedright}
    \begin{center}
        $\begin{array}{ll}
            & \PROCEDURE{\algo}\\
            & \WHILE{\TRUE}\\
            & \BRACE{\IS{x}{\sub{gauss}(0,1 / \sqrt{\beta})}\\
                     \IS{y}{\ranc{0, \pitwo(x)}}\\
                     \IF{y < \pitwofour(x)}\ \BREAK \\
            }\\
            & \OUTPUT{x}\\
            & \ENDPROCEDURE\
        \end{array}$
    \end{center}
\caption{\sub{\algo}. Sampling $\pitwofour$ through the
rejection of Gaussians samples from \Eq{equ:GaussianBell}. }
\label{alg:\algo}
\end{algorithm}

\section{Reversible Markov chains}
\label{sec:reversible_markov_chains}

The probability of rejecting a pebble in \prog{direct-sampling} is
not too high, and a sample of $\pitwofour$ is obtained in a split
second from a sample of $\pitwo$. In real life, however, the difference
between any distribution that we can sample (as $\pitwo$) and the one we want to
sample (as $\pitwofour$) becomes huge, thwarting the direct-sampling approach.
In the alternative Markov-chain sampling, one starts at time $t=0$ with a sample
$x_0$ from a distribution $\pit{0}$ that one knows how to sample. At the next
step, the position $x_1$ samples a distribution $\pit{1}$, and so on.
Introducing the transition matrix $P$ such that $P(x', x)$ represents the
probability to move from a sample $x'$ to
a sample $x$ in one time step, the distribution at time $t+1$ can be expressed
as
\begin{equation}
    \pit{t + 1}(x) = \sum_{x' \in \Omega} \pit{t}(x') P(x', x) \quad \forall x
\in \Omega,
\label{equ:GlobalBalanceTime}
\end{equation}
where the sample space $\Omega$ represents the set of all configurations of the system. 
Markov-chain Monte Carlo requires that, at large $t$, 
$x_t$ samples the distribution $\pit{t \to
\infty}=\pi$. For this to take place, the 
transition matrix $P$ must satisfy,
for all $x \in \Omega$, the global-balance condition,
\begin{equation}
    \pi(x) = \sum_{x' \in \Omega} \pi(x') P(x', x)
    \quad\text{(global balance)},
\label{equ:GlobalBalance}
\end{equation}
which is nothing but the steady-state version of \Eq{equ:GlobalBalanceTime}.
The strategy for sampling $\pi$ implied in
\Eqtwo{equ:GlobalBalanceTime}{equ:GlobalBalance} represents a monumental
investment, as we have to
wait a long time until $\pit{t} \sim \pi$ in order to get a single sample of
$\pi$. It is
not uncommon for this mixing
time to correspond to  weeks or even years of computer time~\cite{Li2022}.

\begin{figure}[htb]
\centering
\includegraphics{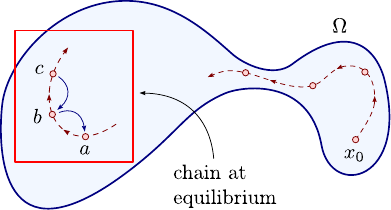}
\caption{Reversible Markov chain. In equilibrium, it satisfies $\prob(a \to b
\to c) = \prob(c \to b \to a)$ for all $a, b, c \in \Omega$.}
\label{fig:DetailedReversible}
\end{figure}

The algorithms in this section are more restrictive than required by
\Eq{equ:GlobalBalance}. They satisfy, for all $x, x' \in \Omega$,
the detailed-balance condition:
\begin{equation}
    \pi(x) P(x, x')  = \pi(x') P(x', x)
    \quad\text{(detailed balance)}.
\label{equ:DetailedBalance}
\end{equation}
It suffices to sum \Eq{equ:DetailedBalance} over all $x' \in \Omega$ (using the
conservation of probabilities $\sum_{x'} P(x, x') = 1$), in order to see that
detailed balance implies global balance.

Detailed-balance algorithms are time-reversible. This means that, at large
$t$ (in equilibrium), any segment of the chain (for example $[a \to b \to c]$
in \fig{fig:DetailedReversible}) at subsequent time steps is sampled with the
same probability $\prob$  as the time-reversed segment. In our example, $\prob(a
\to b \to c)$ is pieced together from the probability $\pi(a)$ to sample $a$
and the transition-matrix  probabilities to move  from $a$ to $b$ and then
from $b$ to $c$,  so that
\begin{multline}
\prob(a \to b \to c) = \underbrace{\pi(a) P(a,b)}_{\pi(b) P(b,a)\ \text{\etc}}
P(b,c)\\
= \pi(c) P(c,b) P(b,a) = \prob(c \to b \to a),
\end{multline}
where we have twice used the detailed-balance condition.
By construction, reversible algorithms thus have no net flows (the flow $a \to 
b \to c$ is cancelled
by the flow $c \to b \to a$), and this points to a very serious restriction
imposed by the detailed-balance condition: they can usually only move around
$\Omega$ diffusively, that is, slowly.

In this section, we will first discuss the seminal reversible algorithm due to
Metropolis et al. (\subsect{subsec:Metropolis}). We will then explore
a variant of the Metropolis algorithm which introduces 
a crucial factorization (\subsect{subsec:FactorizedMetropolis}). We finally 
discuss the consensus principle at the origin of modern developments (\subsect{subsec:Consensus}).

\subsection{The Metropolis chain} 
\label{subsec:Metropolis}

To sample the distribution $\pitwofour$
with a reversible transition matrix $P(x,x')$,
we impose the detailed-balance condition $\pi(x) P(x,x') = \pi(x') P(x',
x)$ for any pair $x$ and $x'$ in $\Omega$. To this end, we may choose
\begin{equation}
\pi(x) P(x,x') \propto \minc{\pi(x), \pi(x')}\quad \text{for $x \neq x'$}.
\label{equ:MetropolisTransitionMatrix}
\end{equation}
The right-hand side of \Eq{equ:MetropolisTransitionMatrix} is symmetric in $x$ 
and $x'$, so that the left-hand side 
must also be symmetric. Therefore, detailed balance is automatically satisfied. 
Dividing both sides by
$\pi(x)$, we arrive at the equation famously proposed by Metropolis et al. in
1953:
\begin{equation}
P^{\text{Met}}(x,x') \propto \minc{1, \fracd{\pi(x')}{ \pi(x)}}
\quad \text{for $x \neq x'$}.
\label{equ:MetropolisFilterRough}
\end{equation}
Let us discuss the difference between a transition matrix and a filter,
in order to render \Eq{equ:MetropolisFilterRough} explicit
and get rid of the proportionality sign.
Indeed, the move from $x$ to $x' \neq x$ proceeds in two steps. It
is first proposed with a symmetric \emph{a priori} probability $\ACAL(x, x')$ 
and then is accepted or rejected with a filter:
\begin{equation*}
\underbrace{P^{\text{Met}}(x,x')}_{\text{transition matrix}} =
\underbrace{\ACAL(x,x')}_{\!\!\!\!\!\text{\emph{a priori}
probability}\!\!\!\!\!\!\!\!\!} \overbrace{\pMet(x,x')}^{\text{Metropolis
filter}}.
\end{equation*}
For the Metropolis algorithm, a
proposed move $x \to x'$ (with $x'\neq x$) is thus accepted with probability
\begin{equation}
\pMet (x,x') = \minc{1, \fracd{\pi(x')}{ \pi(x)}}.
\label{equ:MetropolisFilter}
\end{equation}
If the move $x \to x'$ is rejected, the particle remains at $x$. This
sets the diagonal transition matrix elements $P(x,x)$ and guarantees
that $\sum_{x'} P(x,x')=1$.

\progg{metropolis} implements the symmetric \emph{a priori} probability as a
uniform displacement $\Delta = x' - x$ which is as likely  as $-\Delta$. The
Metropolis filter
is implemented with a uniform random number $\Upsilon$ between $0$ and $1$,
that we refer to as a \quot{pebble}.
For large times $t$, when the initial configuration is forgotten,
the algorithm samples $\pitwofour$. In all the following Markov-chain
algorithm, this large-$t$ condition is silently understood.

\begin{algorithm}
    \newcommand{\algo}{metropolis}
    \captionsetup{margin=0pt,justification=raggedright}
    \begin{center}
        $\begin{array}{ll}
            & \PROCEDURE{\algo}\\
            & \INPUT{x}\ \quad \COMMENT{sample at time $t$}\\
            & \IS{\Delta}{\ranb{-\delta,\delta}}\\
            & \IS{x'}{x + \Delta}\\
            & \IS{\Upsilon}{\ranb{0,1}}\\
            & \IF{\Upsilon < \minc{1, \dfrac{\pitwofour(x')}{\pitwofour(x)}}}\
\IS{x}{x'}\\
            & \OUTPUT{x}\ \quad \COMMENT{sample at time $t+1$}\\
            & \ENDPROCEDURE\
        \end{array}$
    \end{center}
    \caption{\sub{\algo}. Sampling $\pitwofour$ with the Metropolis algorithm.
    }
\label{alg:\algo}
\end{algorithm}

\subsection{Factorizing the Metropolis filter}
\label{subsec:FactorizedMetropolis}
The Metropolis algorithm is really famous, but it is not the end of history.
A modern variant is useful for distributions $\pi$ that factorize:
\begin{equation}
\pi = \pi_a \pi_b \pi_c \cdots\pi_k = \prod_{\xi=a\TO k} \pi_\xi.
\end{equation}
For example, the Boltzmann distribution $\pi= 
\expb{-\beta U}$ takes the above form if 
its  potential $U$ can be written as the sum over pair potentials. 
The Metropolis filter of
\Eq{equ:MetropolisFilter} is then
\begin{multline}
\pMet(x,x') = \minc{1, \fracd{\pi_a(x') \pi_b(x')\cdots\pi_k(x')}
{\pi_a(x) \pi_b(x)\cdots\pi_k(x)}} \\
= \minc{1, \prod_{\xi=a\TO k} \fracd{\pi_\xi(x')}{\pi_\xi(x)}},
\label{equ:FactMetropolisFilter}
\end{multline}
and it is implemented in this way in countless computer programs.
An alternative  to \Eq{equ:FactMetropolisFilter} is the factorized Metropolis
filter~\cite{Michel2014JCP},
\begin{multline}
\pFact(x,x') =
\minc{1, \fracd{\pi_a(x')}
{\pi_a(x)}} \cdots
\minc{1, \fracd{\pi_k(x')}
{\pi_k(x)}} \\
= \prod_{\xi=a\TO k}\minc{1,  \fracd{\pi_\xi(x')}{\pi_\xi(x)}}.
\label{equ:FractorizedGeneral}
\end{multline}
If used naively, it gives lower acceptance probabilities than the
Metropolis filter, but it also satisfies the detailed-balance condition.
Let us prove this for the anharmonic oscillator, where
\begin{equation}
    \pFact_{24}(x, x') = \minc{1, \fracb{\pitwo(x')}{\pitwo(x)}}
    \minc{1, \fracb{\pifour(x')}{\pifour(x)}},
\label{equ:FactorizedTwoFour}
\end{equation}
and where
\begin{multline}
\underline{ \pitwofour(x)} =
 \expb{-\fracc{x^2}{2} - \fracc{x^4}{4}} = \\
 \expb{-\fracc{x^2}{2}} \expb{-\fracc{x^4}{4}} =
\underline{\pitwo(x)\pifour(x)},
\end{multline}
illustrating that a potential that is a sum of terms yields a
Boltzmann distribution that factorizes.
Detailed balance is satisfied because of the following:
\begin{multline}
\underline{
\pitwofour(x) P^{\text{fact}}_{24}(x, x') } \\
\propto\underbrace{ \pitwo(x)  \minc{1, \fracb{\pitwo(x')}{\pitwo(x)}}}_{
\minc{\pitwo(x),\pitwo(x')}:\ x
\Leftrightarrow x'}
    \underbrace{\pifour(x) \minc{1, \fracb{\pifour(x')}{\pifour(x)} }}_{
    \minc{\pifour(x),\pifour(x')}:\ x
\Leftrightarrow x'}\\
    \propto \underline{ \pitwofour(x') P^{\text{fact}}_{24}(x', x)},
\end{multline}
where we have dropped the symmetric \emph{a priori} probability $\ACAL$.
\progg{factor-metropolis} samples $\pitwofour$. It implements the factorized
filter in a way that we will soon discover to be naive.
\begin{algorithm}
    \newcommand{\algo}{factor-metropolis}
    \captionsetup{margin=0pt,justification=raggedright}
    \begin{center}
        $\begin{array}{ll}
                & \PROCEDURE{\algo}\\
                & \INPUT{x}\\
                & \IS{\Delta}{\ranb{-\delta,\delta}}\\
                & \IS{x'}{x + \Delta}\\
                & \IS{\Upsilon}{\ranb{0,1}}\\
                & \IF{\Upsilon < \minc{ 1,\fracd{\pi_2(x')}{\pi_2(x)}} \minc{1, 
                \fracd{\pi_4(x')}{\pi_4(x)}}}\\[6pt]
                & \BRACE{\IS{x}{x'}}\\
                & \OUTPUT{x}\\
                & \ENDPROCEDURE\
          \end{array}$
    \end{center}
    \caption{\sub{\algo}. Sampling $\pitwofour$ naively with the factorized
    Metropolis filter.}
\label{alg:\algo}
\end{algorithm}

\subsection{The consensus principle}
\label{subsec:Consensus}
The factorized Metropolis algorithm will turn out to be particularly powerful,
in the
presence of many factors, even an infinite number of them. This is because of
the
consensus principle, that we now discuss,  and which, in the end, will avoid the
evaluation of the lengthy product in \Eq{equ:FractorizedGeneral}.
\begin{table}[h]
    \begin{center}
        \begin{tabular}{ |c|c|c| }
            \hline
            \backslashbox{Harmonic}{Quartic} & Accept ($p_4$)& Reject ($1-p_4$)
\\
            \hline
            Accept ($p_2$) & $p_2p_4$ \cmark & $p_2 (1-p_4)$ \\
            \hline
            Reject ($1-p_2$) & $(1-p_2) p_4$ & $(1-p_2)(1-p_4)$ \\
            \hline
        \end{tabular}
    \end{center}
    \caption{
Consensus principle in the factorized Metropolis filter. A move $x \to
x'$ that is accepted/rejected independently by the harmonic and the quartic
factor
with probabilities taken from \Eq{equ:Consensus}. The acceptance \quot{by
consensus} reproduces the correct probability of \Eq{equ:FactorizedTwoFour}.}
\label{tab:Consensus}
\end{table}
For the anharmonic oscillator, the consensus principle simply relies on the 
fact that the filter
\begin{equation}
    \pFact_{24}(x, x') = \underbrace{\minc{1,
\fracb{\pitwo(x')}{\pitwo(x)}}}_{\text{$p_2$ (in \tab{tab:Consensus})}}
    \underbrace{\minc{1, \fracb{\pifour(x')}{\pifour(x)}}}_{\text{$p_4$ (in
\tab{tab:Consensus})}}
\label{equ:Consensus}
\end{equation}
is a product $p_2 p_4$ of probabilities that may be interpreted
as independent (see \tab{tab:Consensus}).
This holds although the two factors are evidently correlated and, for example,
$\pitwo$ is small when $\pifour$ is. In \prog{factor-metropolis(patch)},
two independent decisions are taken, one for the harmonic and one for
the quartic factor, and the proposed move is finally accepted only if it is
accepted by both factors. The output is identical to that of \prog{factor-metropolis},
and it again samples the Boltzmann distribution $\pitwofour$.

\begin{algorithm}
    \newcommand{\algo}{factor-metropolis(patch)}
    \captionsetup{margin=0pt,justification=raggedright}
    \begin{center}
        $\begin{array}{ll}
            & \PROCEDURE{\algo}\\
            & \INPUT{x}\\
            & \IS{\Delta}{\ranb{-\delta,\delta}}\\
            & \IS{x'}{x + \Delta}\\
            & \IS{\Upsilon_2}{\ranb{0,1};}\ \IS{\Upsilon_4}{\ranb{0,1}}\\[2pt]
            & \IF{\Upsilon_2 < \minc{ 1,\fracd{\pi_2(x')}{\pi_2(x)}} \AND\
             \Upsilon_4 < \minc{ 1,\fracd{ \pi_4(x')}{\pi_4(x)}}}\\
            & \BRACE{\IS{x}{x'}}\ \quad \COMMENT{move accepted by
            consensus}\\
            & \OUTPUT{x}\\
            & \ENDPROCEDURE\
        \end{array}$
    \end{center}
\caption{\sub{\algo}. Patch of \progn{factor-metropolis}, implementing the
consensus principle.}
\label{alg:\algo}
\end{algorithm}
            
\section{Going beyond reversibility}
\label{sec:NonReversible}

In a tradition that started with the Metropolis algorithm, many decades ago,
Markov chains are normally designed with the quite restrictive 
detailed-balance condition,
although they are only required to satisfy global balance.
In this section, we
illustrate modern attempts to overcome the detailed-balance condition in a
systematic way, within the framework of \quot{lifted} Markov
chains~\cite{Diaconis2000,Chen1999}. Our first
lifted Markov chain, \prog{lifted-metropolis}, holds in fewer than a dozen
lines of code, but is quite intricate
(\subsect{subsec:Lifted}). In recent applications, lifted Markov chains
are often formulated for continuous time. For the anharmonic oscillator, this
gives the \quot{zig-zag} algorithm~\cite{Bierkens2019}, where the particle
moves back and forth as
in molecular dynamics  (as in \prog{isolated-dynamics}), but at fixed velocity.
Newton's equations are not solved, but $\pitwofour$ is still sampled exactly,
and quite magically so (\subsect{subsec:ZigZag}). The decision to
reverse the velocity (from \quot{zig} to \quot{zag})  may again be broken up
into independent decisions of the harmonic and the quartic factors foreshadowing
strategies that have  profoundly impacted real-life sampling approaches 
(\subsect{subsec:ZigZagConsensus}).

\subsection{Lifting the Metropolis chain}
\label{subsec:Lifted}

The Metropolis algorithm, from a position $x$, proposes positive and negative
displacements $\Delta$ for the anharmonic oscillator with symmetric \emph{a
priori} probabilities (see \prog{metropolis}). The filter then imposes that the
net flow vanishes, so there will be as many particles to go from $x$ to $x +
\Delta$ as in the reverse direction, even if, say,  $\pi(x) \ll \pi(x +
\Delta)$.

\begin{figure}[htb]
    \centering
    \includegraphics{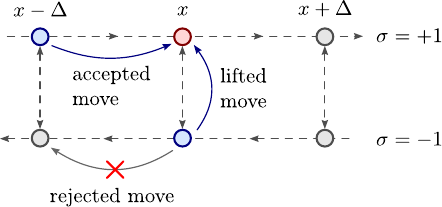}
    \caption{Discretized lifted Metropolis algorithm for the anharmonic
oscillator. The flow into the lifted configuration $\SET{x, +1}$ is
indicated (see \Eq{equ:GlobalLiftedMet}). }
\label{fig:lifted_metropolis}
\end{figure}

We will now break  detailed balance with a non-reversible \quot{lifted} Markov
chain~\cite{Diaconis2000,Chen1999} that only respects global balance, while
having $\pitwofour$ as its stationary distribution. Let us suppose, in a first
step, that the positions $x$ lie on the grid
$\SET{\dots, -2\Delta, -\Delta, 0, \Delta, 2\Delta, \dots}$, with
moves allowed only between nearest neighbors.
Each configuration
$x$ is duplicated (\quot{lifted})  into two configurations, a forward-moving
one $\SET{x, + 1}$, and a backward-moving one
$\SET{x, -1}$. From a lifted configuration $\SET{x,\sigma}$, the lifted
Metropolis algorithm only proposes a forward move if $\sigma=1$, and only a
backward move if $\sigma=-1$. In summary,
\begin{equation*}
\Plift \glb \SET{x,\sigma}, \SET{x+ \sigma \Delta, \sigma} \grb =
\minc{1,
\fracd{\pitwofour(x + \sigma \Delta)}{\pitwofour(x)}},
\end{equation*}
where $\sigma = \pm 1$.
When this move is not accepted by the Metropolis
filter, the algorithm flips the direction and instead moves from
$\SET{x,\sigma}$ to $\SET{x, -\sigma}$:
\begin{equation}
\Plift (\SET{x,\sigma}, \SET{x, -\sigma}) = 1- \minc{1,
\fracd{\pitwofour(x + \sigma \Delta)}{\pitwofour(x)}}.
\end{equation}
This algorithm clearly violates detailed balance as, for example,
\begin{align*}
\Plift(\SET{x, +1},\SET{x + \Delta, +1})& > 0,\\
\Plift(\SET{x + \Delta, +1},\SET{x, +1})&= 0.
\end{align*}
There is thus no backward flow for $\sigma = +1$
and no forward flow for $\sigma = -1$.
On the other hand, the lifted Metropolis algorithm
satisfies the global-balance condition of \Eq{equ:GlobalBalance}
with the \quot{ansatz}
\begin{equation}
\pitwofour^{\text{lift}}(\SET{x, \sigma}) = \half \pitwofour(x)
\quad \text{for $\sigma = \pm 1$}.
\label{equ:AnsatzLifted}
\end{equation}
For example, the flow into the lifted configuration $\SET{x,+1}$ satisfies
\begin{multline}
 \pitwofour(\SET{x,+1}) = \\
 \pitwofour(\SET{x - \Delta,+1})
 \Plift (\SET{x-\Delta, +1},\SET{x, +1}) \\ +
 \pitwofour(\SET{x ,-1}) \Plift (\SET{x, -1},\SET{x, +1}).
\label{equ:GlobalLiftedMet}
\end{multline}
The two contributions on the right-hand side of \eq{equ:GlobalLiftedMet}
correspond on the one hand to the
accepted moves from $\SET{x-\Delta,+1}$, and on the other hand to the lifted
moves from $\SET{x,-1}$, when the move
from $\SET{x,-1}$ towards $\SET{x - \Delta,-1}$ is rejected (see
\fig{fig:lifted_metropolis}).
\Eqq{equ:GlobalLiftedMet} can be transformed into
\begin{multline*}
 \pitwofour(x) =
 \pitwofour(x - \Delta)
\minc{1,
\fracd{\pitwofour(x)}{\pitwofour(x- \Delta)}} + \\
 \pitwofour(x) \gld 1 -
\minc{1,
\fracd{\pitwofour(x-\Delta\sigma)}{\pitwofour(x)}} \grd,
\end{multline*}
which is identically satisfied. We have shown that the lifted Metropolis
algorithm satisfies the global-balance condition for the ansatz of
\eq{equ:AnsatzLifted}, which splits $\pitwofour(x)$ equally between
$\SET{x,+1}$ and $\SET{x,-1}$. The sequence $ \pit{t}$ will
actually converge towards this stationary distribution under very mild
conditions that are satisfied for the anharmonic
oscillator~\cite{Levin2008,Krauth2021eventchain}.

In the lifted Metropolis algorithm, the particle,
starting from $x_0 = 0$,
climbs uphill in  direction $\sigma$ until
a move is rejected by the filter, when it
remains at
its current position but reverses its velocity to $-\sigma$.
The following downhill
moves, again without rejections, are followed by another uphill climb, and so
on, criss-crossing
between the two wings of the potential $\Utwofour$.
\progg{lifted-metropolis} implements a version of the lifted Metropolis
algorithm where the displacements $\Delta$
are sampled from a positive interval.
The algorithm outputs lifted configurations $\SET{x,\sigma}$ of which,
remarkably,
the $x$ positions sample $\pitwofour$.

\begin{algorithm}
    \newcommand{\algo}{lifted-metropolis}
    \captionsetup{margin=0pt,justification=raggedright}
    \begin{center}
        $\begin{array}{ll}
            & \PROCEDURE{\algo}\\
            & \INPUT{\SET{x,\sigma}}\ \quad \COMMENT{lifted sample at time
$t$}\\
            & \IS{\Delta }{\ranb{0,\delta}}\ \quad
            \COMMENT{$\delta > 0$}\\
            & \IS{x'}{x + \sigma \Delta}\ \quad \COMMENT{$x'$ in
            direction $\sigma $ from $x$}\\
            & \IS{\Upsilon}{\ranb{0,1}}\\
            & \IF{\Upsilon < \minc{1, \fracd{\pitwofour(x')}{\pitwofour(x)}}}\ 
            \IS{x}{x'}\\
            & \ELSE\ \IS{\sigma}{-\sigma}\\
            & \OUTPUT{\SET{x, \sigma}}\quad \COMMENT{lifted sample at time
$t+1$}\\
            & \ENDPROCEDURE\
        \end{array}$
    \end{center}
\caption{\sub{\algo}. Non-reversible lifted version of \prog{metropolis}.
The $x$-positions that are output by this program sample $\pitwofour$. }
\label{alg:\algo}
\end{algorithm}

\subsection{From discrete to continuous time}
\label{subsec:ZigZag}

\begin{figure}[htb]
    \centering
    \includegraphics{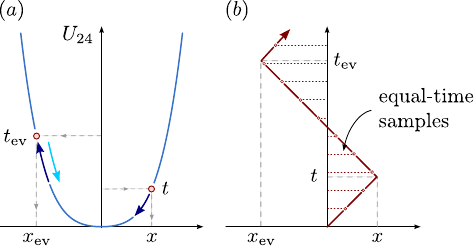}
    \caption{Zig-zag algorithm (continuous-time event-driven
lifted Metropolis chain). \subcap{a} The particle swings about the origin,
turning around at  positions $\xeventtwofour$ (sampled
by \Eq{equ:zig-zag_x}). \subcap{b} Piecewise deterministic constant-velocity
trajectory. Particle positions are sampled at equal time steps.}
\label{fig:zig-zag}
\end{figure}

So far, we have discussed Markov chains that move between configurations indexed
by an integer time $t$, from $x_t $ to $x_{t+1}$. We now consider algorithms in
continuous time (technically speaking, we consider Markov \quot{processes}). For
simplicity, we revisit the lifted Metropolis algorithm with its grid of
positions $\SET{\dots, -2\Delta, -\Delta, 0, \Delta, 2\Delta, \dots}$ and with
its nearest-neighbor moves, but consider the case of small $\Delta$. It is then
appropriate to rescale time such that a displacement $\pm \Delta$ is itself
undertaken
in a time interval $\Delta$. The particle in the anharmonic oscillator thus
moves with unit absolute velocity, whose sense is reversed when there is a
rejection. The downhill moves are all accepted, and even uphill moves
are accepted with a probability close to one.
One may sample the position of the next rejection, rather than
running through the
sequence of individual moves, because an uphill move starting, say, in positive
direction from $x=0$
is accepted with probability $\expc{-\beta \Delta\Utwofour(x=0)}$. 
Likewise, the probability for accepting a whole
sequence of $n$ uphill moves, at subsequent positions $0, \Delta \TO (n-1)
\Delta$, and then rejecting the move $n + 1$, is
\begin{multline}
        \prob (0 \to \xeventtwofour) =\\
        \underbrace{\expa{-\beta
        \Delta\Utwofour(0) \cdots
        \Delta\Utwofour[(n-1)\Delta]}}_{n \text{ accepted moves}} \,
        \underbrace{\glc 1 - \expa{-\beta\Delta\Utwofour(n\Delta)}
        \grc}_{\text{rejection}}  \\
         \to  \beta \expa{-\beta \Utwofour} \ddd\Utwofour.
\label{equ:ExponentialEvent}
\end{multline}
In the small-$\Delta$ limit, the rejection is here expanded to first order, and
$\Delta U $ is replaced by $\ddd U$. In our example of the anharmonic oscillator
starting at $x=0$, all the increments of $\Delta \Utwofour$ up to position $x$
add up to the potential $\Utwofour(x)$. \Eqq{equ:ExponentialEvent} indicates
that the value of $\Utwofour$ at which the velocity is reversed
follows an exponential distribution in $\Utwofour$~\cite{Peters_2012}.
As an exponential random number can be obtained as a logarithm of a
uniform random number (see~\cite[Sec. 1.2.4]{SMAC}), this yields
\begin{equation}
    \Utwofour(\xeventtwofour) = - \beta^{-1} \, \loga{\ranb{0, 1}}.
\label{equ:zig-zag_U}
\end{equation}
Inverting $\Utwofour(\xeventtwofour) =
\xeventtwofour^2/2 +
\xeventtwofour^4/4 $,
this results in
\begin{equation}
    \xeventtwofour =  \sigma \sqrt{-1 + \sqrt{1 - 4 \beta^{-1} \, \loga{\ranb{0,
1}}}}.
\label{equ:zig-zag_x}
\end{equation}
To sample the Boltzmann distribution $\pitwofour$, it now
suffices to sample the turning points $\xeventtwofour$ of the
constant-velocity motion, alternatingly on the negative and positive branches
of the potential,
and then to sample
the particle positions at equal time steps, as implemented
in \prog{zig-zag}. This event-driven
continuous-time algorithm samples the Boltzmann distribution 
$\pitwofour$ (see 
\fig{fig:zig-zag}).
The event-driven version of \progn{lifted-metropolis}
exists also for fixed, finite $\Delta$,
and it is often classified as \quot{faster-than-the-clock}
(see~\cite[Sec. 7.1.1]{SMAC}).

\begin{algorithm}
    \newcommand{\algo}{zig-zag}
    \captionsetup{margin=0pt,justification=raggedright}
    \begin{center}
        $\begin{array}{ll}
            & \PROCEDURE{\algo}\\
            & \INPUT{\SET{x, \sigma}, t}\ \quad \COMMENT{lifted sample
                       with $\sigma x \le 0$ }\\[2pt]
            & \IS{\xeventtwofour}{\sigma\sqrt{-1 + \sqrt{1 - 4 \beta^{-1} \,
\loga{\ranb{0,
            1}}}}}\ \quad
\COMMENT{see \Eq{equ:zig-zag_x}}\\
            & \IS{\tevent}{t + |\xeventtwofour - x|}\\
            & \FOR{t^* = \text{int}(t) + 1 \TO
\text{int}(\tevent)}\\
            & \BRACE{\PRINT{x + \sigma (t^* - t)}\\
                    }\ \quad \COMMENT{equal-time samples}\\
            &
            \IS{x}{\xeventtwofour};\
            \IS{\sigma}{-\sigma};\
            \IS{t}{\tevent}\ \quad \COMMENT{\quot{zig-zag}}\\
            & \OUTPUT{\SET{x, \sigma}, t}\\
            & \ENDPROCEDURE\
        \end{array}$
    \end{center}
\caption{\sub{\algo}. Continuous-time version of \prog{lifted-metropolis}
using an event-driven formulation. The $x$-positions output by the 
\sub{print} statement sample $\pitwofour$.}
\label{alg:\algo}
\end{algorithm}
        
\subsection{Extending the consensus principle}
\label{subsec:ZigZagConsensus}

\begin{figure}[htb]
    \centering
    \includegraphics{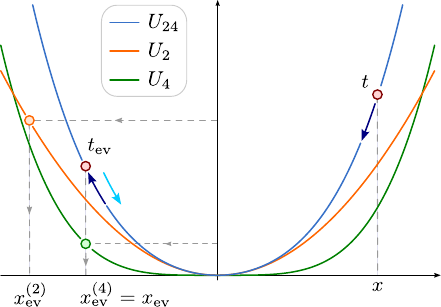}
    \caption{Factorized zig-zag algorithm. Starting from $x$ (here with
$\sigma = -1$, the next event is
    given by the earliest event between $\xeventtwo$ and $\xeventfour$ (here,
by $\xeventfour = \xeventtwofour$).}
\label{fig:FactorizedZigZag}
\end{figure}

We now replace the Metropolis filter in \prog{zig-zag} (contained in the
formula for $\xeventtwofour$) by the factorized
Metropolis filter, and then use the consensus principle. Starting again at
$x=0$,
the particle now climbs up
one hill
for the harmonic factor and one for the quartic factor (see
\fig{fig:FactorizedZigZag}). For each factor, we can redo the argument of
\Eq{equ:ExponentialEvent}, with $\Utwo$ or $\Ufour$ instead of
$\Utwofour$. In analogy with \eqtwo{equ:zig-zag_U}{equ:zig-zag_x},
we can thus sample two \quot{candidate}
events,
\begin{align}
    & \xeventtwo = \sigma \sqrt{-2 \beta^{-1} \, \loga{\ranb{0, 1}}},
\label{equ:zig-zag_consensus_x2} \\[2pt]
    & \xeventfour = \sigma \sqrt[4]{-4 \beta^{-1} \, \loga{\ranb{0, 1}}},
\label{equ:zig-zag_consensus_x4}
\end{align}
with two independent random numbers. The consensus of the two factors is broken
by the candidate event that comes first,
\begin{equation}
    \xeventtwofour = \sigma \minb{|\xeventtwo|, |\xeventfour|},
\label{equ:zig-zag_consensus_x}
\end{equation}
when the velocity must be reversed. We may
again collect positions $x$ at equal time steps. This is implemented in
\prog{factor-zig-zag}, which samples the Boltzmann distribution $\pitwofour$.

\begin{algorithm}
    \newcommand{\algo}{factor-zig-zag}
    \captionsetup{margin=0pt,justification=raggedright}
    \begin{center}
        $\begin{array}{ll}
            & \PROCEDURE{\algo}\\
            & \INPUT{\SET{x, \sigma}, t}\ \quad \COMMENT{lifted sample
            with $\sigma x \le 0$}\\[2pt]
            & \IS{\xeventtwo}{\sigma \sqrt{- 2 \beta^{-1} \, \loga{\ranb{0,
1}}}}\ \quad
\COMMENT{see \Eq{equ:zig-zag_consensus_x2}}\\[2pt]
            & \IS{\xeventfour}{\sigma \sqrt[4]{- 4 \beta^{-1} \, \loga{\ranb{0,
1}}}}\ \quad
\COMMENT{see \Eq{equ:zig-zag_consensus_x4}}\\
            & \IS{\xeventtwofour}{\sigma \minb{|\xeventtwo|, |\xeventfour|}}\\
            & \IS{\tevent}{t + |\xeventtwofour - x|}\\
            & \FOR{t^* = \text{int}(t) + 1 \TO \text{int}(\tevent) } \\
            & \BRACE{\PRINT{x + \sigma (t^* - t)}\\
                    }\ \quad \COMMENT{sample of $\pitwofour$}\\
           & \IS{x}{\xeventtwofour};\
            \IS{\sigma}{-\sigma};\
            \IS{t}{\tevent}\ \quad \COMMENT{\quot{zig-zag}}\\
            & \OUTPUT{\SET{x, \sigma}, t}\\
            & \ENDPROCEDURE\
         \end{array}$
    \end{center}
\caption{\sub{\algo}. Factorized zig-zag algorithm accepting moves in direction
$\sigma$ until the consensus of the harmonic and the quartic factor is broken
at position $\xeventtwofour$.}
\label{alg:\algo}
\end{algorithm}

\section{Thinning: or, avoiding evaluation}
\label{sec:Thinning}

In molecular-dynamics algorithms such as \progn{thermostat-dynamics}, forces
must be computed precisely in order to keep the trajectory on track. In
contrast, Monte Carlo algorithms are decision problems where proposed moves must
be accepted with a filter, for example the Metropolis filter $\min[1,
\expb{-\beta \Delta U}]$. As we discuss in this section, one can often base the
accept/reject decision on a bounding potential $\Ubound$, and thus avoid
computing $U$, $\Delta U$, and their exponentials
(\subsect{subsec:BoundingPotential}). In the continuous-time setting, one
simply evaluates the derivative of the bounding potential and of the
potential $U$, in order to eliminate all bias due to the bounding
(\sect{subsec:ContinuousZigZag}).

Combining this so-called
\quot{thinning} approach~\cite{LewisShedler1979} with the factorization, we may,
in the anharmonic oscillator, base our decision to accept moves on the consensus
of harmonic and quartic bounding potentials. At the end, we will set up a Monte
Carlo
algorithm that evaluates a single factor potential, and only at the position
where the proposed move is rejected by the bounding potential of that same
factor (\subsect{subsec:ThinningConsensus}).
In the companion paper~\cite{Tartero2023b}, we
generalize this approach to real-life simulations of particles with long-range
interactions that sample the Boltzmann distribution $\expb{-\beta U}$ without
ever evaluating $U$.

\subsection{Introducing the bounding potential}
\label{subsec:BoundingPotential}

\begin{figure}[htb]
    \centering
    \includegraphics{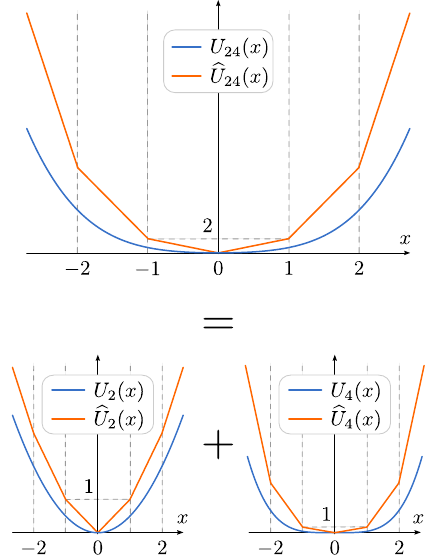}
    \caption{Anharmonic bounding potential $\Uboundtwofour$
        and its harmonic and quartic constituents $\Uboundtwo$ and 
        $\Uboundfour$.}
    \label{fig:BoundingPotential}
\end{figure}

We say that $\Ubound$ is a bounding potential of a potential $U$ if, for any
pair of configurations $x$ and $x'$, it satisfies
\begin{equation}
  \minb{1, \expa{- \beta \Delta \Ubound}} \leq \minb{1,\expa{- \beta \Delta U}}
\quad
  \forall \, x, x' \in \Omega,
\label{equ:BoundingDefinition}
\end{equation}
where $\Delta \Ubound = \Ubound(x') - \Ubound(x)$ and $\Delta U = U(x') -
U(x)$. This requires
$\diff \Ubound/ \diff x $ and
$\diff U/ \diff x $ to have the same sign everywhere, with
$|\diff \Ubound/ \diff x |\ge |\diff U/ \diff x| $.
Concretely, we define the harmonic and
quartic bounding potentials as
\begin{equation*}
    \begin{split}
        & \Uboundtwo(n) =
        \begin{cases}
            0 & \text{if $n=0$}\\
            \Uboundtwo(|n|-1) + |n| & \text{if $n \in \ZZ \backslash
            \{0\}$}
        \end{cases}
        ,\\[4pt]
        & \Uboundfour(n) =
        \begin{cases}
            0 & \text{if $n=0$}\\
            \Uboundfour(|n|-1) + |n^3| & \text{if $n \in \ZZ \backslash
            \{0\}$}.
        \end{cases}
    \end{split}
\end{equation*}
These definitions are extended to non-integer arguments $x$ through linear
interpolation. The anharmonic bounding potential is then defined as
$\Uboundtwofour(x) = \Uboundtwo(x) + \Uboundfour(x)$ (see
\fig{fig:BoundingPotential}).

\begin{figure}[htb]
    \centering
    \includegraphics{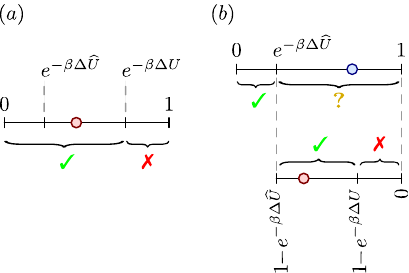}
    \caption{Single-pebble and two-pebble decisions in the Metropolis algorithm.
    \subcap{a} A single pebble $\Upsilon$ illustrating that
    acceptance with respect to the bounding potential
    implies acceptance with respect to $U$.
    \subcap{b} A first pebble $\Upsilon_1$ takes a decision with respect to
    the bounding potential. In case of rejection, a second pebble
    $\Upsilon_2$ definitely decides on the move.}
    \label{fig:OnePebbleTwoPebbles}
\end{figure}

A bounding potential can simplify the decision to accept a move as, evidently, a
pebble  $0 < \Upsilon< 1$ that falls below $\expba{-\beta \Delta \Ubound}$ also
falls below $\expba{-\beta \Delta U}$ (see \subfig{fig:OnePebbleTwoPebbles}{a}).
In the remaining algorithms of this paper,
we rather use a two-pebble strategy for the decision to accept or
reject a move.
A first pebble $0< \Upsilon_1 < 1$ then decides whether a move is
accepted with respect to the bounding potential. Otherwise (if $\Upsilon_1$
rejects the move), we use a second pebble $\Upsilon_2$ to decide whether
the first-pebble rejection with respect to $\Ubound$ stands with respect to $U$
(see \subfig{fig:OnePebbleTwoPebbles}{b}). A rescaling, with $0 <
\Upsilon_2<1$, allows us to definitely reject the move if
\begin{equation}
\Upsilon_2 < \fracb{1 - \expa{- \beta \Delta U}}{1
    - \expa{-\beta \Delta \Ubound}}.
\label{equ:thinning_P2}
\end{equation}
The two-pebble bounding-potential algorithm is implemented in
\prog{bounded-lifted} for the anharmonic oscillator.
It again samples the
Boltzmann distribution $\pitwofour$.
\begin{algorithm}
    \newcommand{\algo}{bounded-lifted}
    \captionsetup{margin=0pt,justification=raggedright}
    \begin{center}
        $\begin{array}{ll}
             & \PROCEDURE{\algo}\\
             & \INPUT{\SET{x, \sigma}}\ \quad \COMMENT{lifted sample at time
$t$}\\
             & \IS{\Delta}{\ranb{0,\delta}}\ \quad \COMMENT{$\delta > 0$}\\
             & \IS{x'}{x + \sigma \, \Delta};\ \IS{\Upsilon_1}{\ranb{0,1};}\\
             & \IF{\Upsilon_1 < \minb{1, \expa{-\beta \Delta \Uboundtwofour}}}\\
             & \BRACE{\IS{x}{x'}}\\
             & \ELSE\\

             & \BRACE{
                   \IS{\Upsilon_2}{\ranb{0,1}}\\
                   \IF{ \Upsilon_2 > \fracd{1 - \expa{-\beta \Delta \Utwofour} }
                     {1 - \expa{-\beta \Delta \Uboundtwofour}}}\ \IS{x}{x'}\\
                 \ELSE\ \IS{\sigma}{-\sigma} \\
             }\\
             & \OUTPUT{\SET{x, \sigma}}\quad \COMMENT{lifted sample at time
$t+1$}\\
             & \ENDPROCEDURE\
         \end{array}$
     \end{center}
\caption{\sub{\algo}. Discrete-time bounded-lifted Metropolis algorithm
using two-pebble decisions. The second pebble is used
and the true potential $\Utwofour$ is evaluated only after a
first-pebble rejection with respect to the bounding potential $\Uboundtwofour$.}
\label{alg:\algo}
\end{algorithm}
        
\subsection{Continuous-time thinning}
\label{subsec:ContinuousZigZag}

\begin{figure}[htb]
    \centering
    \includegraphics{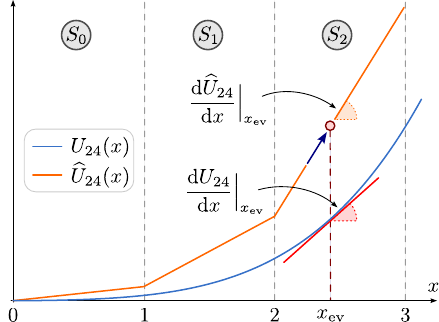}
    \caption{Continuous-time version of the bounded-lifted Metropolis algorithm.
        The proposed event $\xeventtwofour$ is confirmed by comparing
        the derivatives of the true potential $\Utwofour$ and  the
        bounding potential $\Uboundtwofour$ (see \Eq{equ:ThinningContinuous}). }
    
    \label{fig:bounded-zig-zag}
\end{figure}

The bounded-lifted Metropolis algorithm, \prog{bounded-lifted}, generalizes to
continuous time.
In the anharmonic oscillator,
we first consider
$\sigma = +1$ and positive $x$ between $n$ and $n+1$,
where the decision of \Eq{equ:thinning_P2}, for
the second pebble, turns into
\begin{equation}
\Upsilon_2 <
 \fracb{1 - \expa{- \beta \Delta \Utwofour}}{1
    - \expa{-\beta \Delta \Uboundtwofour}}
    \rightarrow
\Upsilon_2 <
        \fracb{\fraca{\diff \Utwofour}{\diff
        x}}{\fraca{\diff \Uboundtwofour}{\diff x}}.
\label{equ:ThinningContinuous}
\end{equation}
The piecewise linear anharmonic bounding potential $\Uboundtwofour$ simplifies
the event-driven formulation. Rather than to walk up the anharmonic potential
until the  change of potential satisfies $\Delta \Utwofour = - \beta^{-1}
\loga{\ranb{0, 1}}$ (see \Eq{equ:zig-zag_U} and \fig{fig:zig-zag}), we
now run up a bounding potential of constant slope
$\qtwofourhat$ with
\begin{equation}
    \qtwofourhat = \fracb{\diff}{\diff x} \Uboundtwofour(x) \Big \vert_{x \in
S_n}
= n +
    1 + (n + 1)^3,
\label{equ:bounded_zig-zag_lambda}
\end{equation}
where $S_n =[n, n + 1)$ and $n \in \mathbb{N}$.
The change in potential $\Delta \Uboundtwofour(x)  = - \beta^{-1} \loga{\ranb{0,
1}}$ then translates into the advance of the  position as
\begin{equation}
     \xeventtwofour = x_0 + \glb \beta \qtwofourhat \grb ^ {-1} \loga{\ranb{0,
1}}.
\label{equ:bounded_zig-zag_x}
\end{equation}
The event rate $\beta \qtwofourhat$ is constant in the sector $S_n$, but
if $\xeventtwofour$ falls outside of $S_n$, it is invalid. In this case, a
\quot{boundary event} is triggered, and the particle is placed at the right
boundary of $S_n$,  without changing the direction $\sigma$. Otherwise (if
$\xeventtwofour \in S_n$), the direction $\sigma$ is reversed if the condition
on the pebble $\Upsilon_2$ in \Eq{equ:ThinningContinuous} is satisfied (see
\fig{fig:bounded-zig-zag}).

Our description of the continuous-time bounded-lifted Metropolis algorithm was
for the case $\sigma=1$, that is, for a pebble that climbs up the $x>0$ branch
of the potential. The general case is implemented in \prog{bounded-zig-zag}, and
it again samples the Boltzmann distribution $\pitwofour$.

\begin{algorithm}
    \newcommand{\algo}{bounded-zig-zag}
    \captionsetup{margin=0pt,justification=raggedright}
    \begin{center}
        $\begin{array}{ll}
            & \PROCEDURE{\algo}\\
            & \INPUT{\SET{x, \sigma}, t }\ \quad \COMMENT{lifted sample}\\
            & \IF{\sigma x < 0}\ \IS{x_0}{0};\ \ELSE\ \IS{x_0}{x}\ \quad
            \COMMENT{starting point}\\
            & \IS{n}{\text{int}(|x_0|)};\ \IS{\qtwofourhat}{n + 1 + (n +1)^3};\ 
            \IS{\stilde}{\sigma}\\
            & \IS{\xeventtwofour}{x_0 + \sigma \glc - (\beta \qtwofourhat)^{-1}
                  \log \ranb{0,1} \grc }\ \quad \COMMENT{see
                     \Eq{equ:bounded_zig-zag_x}}\\
            & \IF{|\xeventtwofour | > n + 1}\\
            & \BRACE{
            \IS{\xeventtwofour}{\sigma (n+1)}
            }\\ 
            & \ELIF{\ranb{0,1} < |\xeventtwofour + \xeventtwofour^3| /
\qhat} \\
            & \BRACE{\IS{\stilde}{-\sigma}}\\
            & \IS{\tevent}{t + |\xeventtwofour - x|}\\
            & \FOR{t^* = \text{int}(t) + 1 \TO \text{int}(\tevent)}\\
            & \BRACE{\PRINT{x + \sigma (t^* - t)}}\ \quad
            \COMMENT{equal-time samples}\\
            & \IS{x}{\xeventtwofour};\ \IS{\sigma}{\stilde};\
            \IS{t}{\tevent}\ \quad \COMMENT{\quot{zig-zag}}\\
            & \OUTPUT{\SET{x, \sigma}, t}\\
            & \ENDPROCEDURE\
        \end{array}$
    \end{center}
    \caption{\sub{\algo}. Continuous-time bounded-lifted Metropolis algorithm.
    It need not invert the potential $\Utwofour$
    (compare with \Eq{equ:zig-zag_x}), foreshadowing the use of
    bounding potentials in real-world applications.}
\label{alg:\algo}
\end{algorithm}

\subsection{Thinning with consensus}
\label{subsec:ThinningConsensus}

\progg{bounded-zig-zag} avoids the inversion in \Eq{equ:zig-zag_x} of the
potential $\Utwofour$, and only evaluates the derivative $\fraca{\diff
\Utwofour}{\diff x}$ at $x = \xeventtwofourhat$. At the
end of our journey through advanced Markov chain Monte Carlo sampling, we
combine the consensus
principle underlying factorization with that of thinned, lifted Metropolis
chains and sample $\pitwofour = \expb{-\beta \Utwofour}$ without ever
evaluating the potential $\Utwofour$ nor its derivative. The use of bounding
potentials
generalizes to applications in particle systems with long-range interactions.
In the anharmonic oscillator, we illustrate the basic
idea~\cite{KapferKrauth2016} with the harmonic and quartic factor potentials
$\Utwo$ and $\Ufour$ and their bounding potentials $\Uboundtwo$ and
$\Uboundfour$.

With factorization, two candidate events $\xeventtwo$ and $\xeventfour$ can
be sampled by means of
\Eq{equ:bounded_zig-zag_x}, with bounding event rates $\beta (n+1)$ and
$\beta (n+1)^3$, respectively. When both events fall outside the sector
$S_n$ where the
bounding rates are valid, a boundary event is triggered. Otherwise, the
earliest candidate event $\xeventtwofour \in S_n$ (either $\xeventtwo$ or
$\xeventfour$) is confirmed
with one of the probabilities
\begin{equation*}
    \fracb{\fraca{\diff \Utwo}{\diff x}}{\fraca{\diff \Uboundtwo}{\diff x}} 
     \Bigg \vert_{\xeventtwofour = \xeventtwo} \!\!\!\!\!\!\!=
    \fracb{\xeventtwofour}{n+1}, \quad \fracb{\fraca{\diff \Ufour}{\diff
    x}}{\fraca{\diff 
    \Uboundfour}{\diff x}} \Bigg \vert_{\xeventtwofour =
\xeventfour}\!\!\!\!\!\!\! =
    \fracb{\xeventtwofour^3}{(n+1)^3}.
\end{equation*}
This bounded-lifted, and in addition factorized, Metropolis algorithm, largely
analogous to \progn{bounded-zig-zag}, is implemented in
\prog{bounded-factor-zig-zag}. Remarkably, it evaluates the derivative of only
one factor potential. Most of the decisional burden is carried by the bounding
potentials, for example which factor to choose for the next event. The
decision-problem footprint of Monte Carlo algorithms thus appears clearly, as
there are different ways to reach a statistically correct decision. In molecular
dynamics, in contrast, only a single Newtonian trajectory exists.

\begin{algorithm}
    \newcommand{\algo}{bounded-factor-zig-zag}
    \captionsetup{margin=0pt,justification=raggedright}
    \begin{center}
        $\begin{array}{ll}
            & \PROCEDURE{\algo}\\
            & \INPUT{\SET{x, \sigma}, t }\ \quad \COMMENT{lifted sample at time 
            $t$} \\
            & \IF{\sigma x < 0}\ \IS{x_0}{0};\ \ELSE\ \IS{x_0}{x}\ \quad 
            \COMMENT{starting point}\\
            & \IS{n}{\text{int}(|x_0|)};\ \IS{\qtwohat}{n + 1};\
                        \IS{\qfourhat}{(n +1)^3};\ \IS{\stilde}{\sigma}\\[2pt]
            & \IS{\xeventtwo}{x_0 + \sigma \glc - (\beta \qtwohat)^{-1}
                  \log \ranb{0,1} \grc }\ \quad \COMMENT{see
                     \Eq{equ:bounded_zig-zag_x}}\\[6pt]
            & \IS{\xeventfour}{x_0 + \sigma \glc - (\beta \qfourhat)^{-1}
                  \log \ranb{0,1} \grc }\ \quad \COMMENT{see
                     \Eq{equ:bounded_zig-zag_x}}\\[4pt]
            & \IF{\min(|\xeventtwo|,|\xeventfour|) > n + 1}
\IS{\xeventtwofour}{\sigma (n+1)}\\[2pt]
            & \ELIF{|\xeventtwo| < |\xeventfour| } \\
            & \BRACE{
                     \IS{\xeventtwofour}{\xeventtwo};\
                     \IF{\ranb{0,1} <
                     \fraca{|\xeventtwofour|}{\qtwohat}}\ \stilde = -\sigma
            }\\
            & \ELSE\\
            & \BRACE{
                 \IS{\xeventtwofour}{\xeventfour};\
                 \IF{\ranb{0,1} <
                 \fraca{|\xeventtwofour|^3}{\qfourhat}}\ \stilde = -\sigma
            }\\
            & \IS{\tevent}{t + |\xeventtwofour - x|}\\
            & \FOR{t^* = \text{int}(t) + 1 \TO \text{int}(\tevent)}\\
            & \BRACE{\PRINT{x + \sigma (t^* - t)}}\ \quad \COMMENT{
               equal-time samples}\\
            & \IS{x}{\xeventtwofour};\ \IS{\sigma}{\stilde};\
            \IS{t}{\tevent}\ \quad \COMMENT{\quot{zig-zag}}\\
            & \OUTPUT{\SET{x, \sigma}, t}\\
            & \ENDPROCEDURE\
        \end{array}$
    \end{center}
    \caption{\sub{\algo}. Factorized version of \progn{bounded-zig-zag}, with
    one candidate event for each factor (see patch). For each event, only one
    factor derivative is evaluated.}
\label{alg:\algo}
\end{algorithm}

\begin{table}[h]
    \begin{center}
        \begin{tabular}{ |c|c|c| }
            \hline
            \backslashbox{Harmonic}{Quartic} & Accept ($1 - \beta \qfourhat
\diff t $)&
Reject ($\beta \qfourhat \diff t$) \\
            \hline
            Accept ($1 - \beta \qtwohat \diff t$) & $1 - \beta(\qtwohat +
\qfourhat) \diff t$
& $\beta \qfourhat \diff t$ \\
            \hline
            Reject ($\beta \qtwohat \diff t $) & $\beta \qtwohat \diff t $
& $0 $ \\
            \hline
        \end{tabular}
    \end{center}
    \caption{
Consensus probabilities of \tab{tab:Consensus} for \prog{bounded-factor-zig-zag}
and its patch. The total event rate (the total rate of rejection by
consensus) is the sum of the factor event rates, as terms of order $(\diff
t)^2$ drop out. }
\label{tab:Consensus2}
\end{table}
\progg{bounded-factor-zig-zag}  samples as many candidate events as there are
factors (in our case, $\xeventtwohat$ and
$\xeventfourhat$ for the harmonic and quartic factors), thus adopting a strategy
that runs into trouble when there are too many factors.
A patch of \progn{bounded-factor-zig-zag} illustrates, in a nutshell,
how factors can be bundled
in the continuous-time setting, where the total event rate is the
sum of the individual factor
rates (see \tab{tab:Consensus2}). In the anharmonic oscillator,
the
total bounding event rate is the sum of the harmonic and the
quartic bounding rates,
giving us the next event with a single random number. It then remains to decide
whether this event is a harmonic-bounding or a quartic-bounding event, as
implemented in
\prog{bounded-factor-zig-zag(patch)}. Even for a large number of factors,
we can take this decision in a few steps, using the famous Walker
algorithm~\cite{Walker1977AnEfficientMethod}.
It is this very program that is used in state-of-the-art
programs to handle millions of factors in constant
time~\cite{KapferKrauth2016}, as we will further
discuss in the companion paper~\cite{Tartero2023b}.

\begin{algorithm}
    \newcommand{\algo}{bounded-factor-zig-zag(patch)}
    \captionsetup{margin=0pt,justification=raggedright}
    \begin{center}
        $\begin{array}{ll}
            & \PROCEDURE{\algo}\\
            & \INPUT{\SET{x, \sigma}, t }\ \quad \COMMENT{lifted sample at time
            $t$} \\
            & \IF{\sigma x < 0}\ \IS{x_0}{0};\ \ELSE\ \IS{x_0}{x}\ \quad
            \COMMENT{starting point}\\
            & \IS{n}{\text{int}(|x_0|)}\\
            & \IS{\qtwohat}{n + 1};\
                        \IS{\qfourhat}{(n +1)^3};\
                        \IS{\qtwofourhat}{\qtwohat + \qfourhat};\ 
                        \IS{\stilde}{\sigma}\\[2pt]
            & \IS{\xeventtwofour}{x_0 + \sigma \glc - (\beta \qtwofourhat)^{-1}
                  \log \ranb{0,1} \grc }\ \quad \COMMENT{see
                     \Eq{equ:bounded_zig-zag_x}}\\[6pt]
            & \IF{|\xeventtwofour|  > n + 1} \IS{\xeventtwofour}{\sigma 
            (n+1)}\\[2pt]
            & \ELIF{\ranb{0, \qtwofourhat} < \qtwohat} \\
            & \BRACE{
                \IF{\ranb{0,1} < \fraca{|\xeventtwofour|}{\qtwohat}}\
                \stilde = -\sigma
            }\\ 
            & \ELSE \\
            & \BRACE{
                 \IF{\ranb{0,1} < \fraca{|\xeventtwofour|^3}{\qfourhat}}\
                 \stilde = -\sigma
            }\\
            & \IS{\tevent}{t + |\xeventtwofour - x|}\\
            & \FOR{t^* = \text{int}(t) + 1 \TO \text{int}(\tevent)}\\
            & \BRACE{\PRINT{x + \sigma (t^* - t)}}\ \quad \COMMENT{
               equal-time samples}\\
            & \IS{x}{\xeventtwofour};\ \IS{\sigma}{\stilde};\
            \IS{t}{\tevent}\ \quad \COMMENT{\quot{zig-zag}}\\
            & \OUTPUT{\SET{x, \sigma}, t}\\
            & \ENDPROCEDURE\
        \end{array}$
    \end{center}
    \caption{\sub{\algo}. Patch of \progn{bounded-factor-zig-zag}
illustrating the bundling of two factors into a single candidate event. }
\label{alg:\algo}
\end{algorithm}

\section{Conclusion}
In this paper, we have introduced to
a number of modern developments in Monte Carlo sampling that
go much beyond direct sampling and the Metropolis algorithm.
New Monte Carlo algorithms build on notions such as factorization,
non-reversibility and thinning. They increasingly find
applications in physics and other sciences.
The severely stripped-down one-dimensional anharmonic oscillator has
hopefully allowed us to lay bare the foundations of these non-trivial
theoretical developments. In a first step, we have
concentrated
on the correctness of the sampling algorithms. Questions of
efficiency will be the subject of the companion paper~\cite{Tartero2023b}.

\section*{Acknowledgments}
We thank K. J. Wiese for helpful discussions. We thank the mathematical research
institute MATRIX in Australia where part of this research was performed.

\appendix

\section{Mathematical details}
\label{app:MathDetails}
In this appendix, we present some mathematical details that, for the sake 
of conciseness, were omitted in the main text.

As stated in \Eq{equ:landau_distr}, the period of the isolated anharmonic 
oscillator at energy $E$ is
\begin{equation}
    \Period(E) = 4 \sqrt{\fracb{2}{1 + \sqrt{1 + 4 E}}} \, K \glb
    \fracb{1 - \sqrt{1 + 4 E}}{1 + \sqrt{1 + 4 E}}
    \grb,
\label{equ:TauElliptic}
\end{equation}
where $K$ is the complete elliptic integral of the first kind. This non-trivial
integral follows from the theory of elliptic functions (see $\eg$~\cite[Ch.
19]{NIST} for a discussion on the subject). It can be obtained
indirectly using the \texttt{Integrate} function of the Mathematica
software, as illustrated in a Mathematica notebook file made
available in the software package (see \app{app:Software}).
For $E \to 0$,
the amplitude $\xmax$ of the oscillation is small. Consequently, the anharmonic 
potential $\Utwofour(x)$ of \Eq{equ:potential} can be safely 
replaced with the harmonic one in this regime:
\begin{equation}
    \Utwofour(x) \sim \Utwo(x) = \fraca{x^2}{2} \quad \text{for } \xmax \to 
    0.
\end{equation}
Indeed, expanding   $\Period(E)$ in \Eq{equ:TauElliptic} about
$E = 0$, 
we obtain
\begin{equation*}
    \Period(E) = 2 \pi - \fracb{3 \pi}{2} E + \OCAL(E^2).
\end{equation*}
(see \app{app:Software} for
a Mathematica notebook file using the \texttt{Series} function).

For small $E$, the period of the anharmonic
oscillator coincides with that of the harmonic one, $\Period = 2 \pi$, since 
the quartic term in the potential is negligible for $|x| \ll 1$. On the other
hand, for large $E$,
the quartic term dominates:
\begin{equation*}
    \Utwofour(x) \sim \Ufour(x) = \fraca{x^4}{4} \quad \text{for } \xmax \gg 0.
\end{equation*}
In this case, expanding $\Period$ for large $E$,
we have
\begin{equation*}
    \tau(E) = \fracb{\sqrt{\pi} \, \Gamma(1/4)}{\Gamma(3/4)} E^{-1/4} + 
    \OCAL(E^{-3/4}),
\end{equation*}
where $\Gamma$ denotes the Euler gamma function (see again \app{app:Software}
for the corresponding Mathematica notebook file).
The dominant term of the
above expression 
coincides with the period of the quartic oscillator, computed
using the equivalent of \Eq{equ:landau_distr}, with amplitude
$\xmax = (4E)^{1/4}$.

Finally, the partition function $Z(\beta)$ of the harmonic oscillator in 
\Eq{equ:PartitionAnharmonic} can be easily 
computed by means of the Mathematica \texttt{Integrate} function.

\section{Computer programs, Mathematica notebook files}
\label{app:Software}

The present paper is accompanied by the \MCMCNutshell software package, which is
published as an open-source project under the GNU GPLv3 license. \MCMCNutshell
is
available on GitHub as part of the \texttt{JeLLyFysh} organization~\footnote{The
url of the repository is
\url{https://github.com/jellyfysh/MCMCNutshell}.}.
The
package contains Python implementations for $\beta=1$ of the algorithms that
were discussed
here and that were used to produce the results of \tab{tab:NumericalResults}.
It also contains the Mathematica Notebook files discussed in
\app{app:MathDetails}.

\section{Numerical tests}
\label{app:Tests}

Except for \prog{isolated-dynamics}, the eleven Monte Carlo algorithms and
one molecular-dynamics algorithm all
sample the Boltzmann distribution $\pitwofour$
of \Eq{equ:BoltzmannDistribution}. To check the correctness of our 
implementations,
we fixed an arbitrary non-zero value of $\xbar = 0.63$ for $\beta=1$,
computed for each
algorithm the empirical probability with which  the samples $x $ satisfy $x <
\xbar$, and compared it with the exact result:
\begin{equation}
    \prob(x < 0.63) = Z^{-1} \int_{-\infty}^{0.63} \pitwofour (x') \diff x' =
0.8030245.
\label{equ:Theoretical063}
\end{equation}
Single-standard-deviation error bars
were obtained from the  bunching method~\cite[Sec. 1.3.5]{SMAC}, except
for \prog{direct-sampling}, where we performed a standard Gaussian analysis.
For all twelve algorithms, results are consistent with \Eq{equ:Theoretical063} 
within three standard deviations
(see \tab{tab:NumericalResults}).
\begin{table}[h]
    \begin{center}
        \begin{tabular}{ c|c }
            Algorithm & $\prob(x < 0.63)$ \\
            \hline
            \progtest{thermostat-dynamics} & $ 0.8038 \pm 0.0025 $ \\
            \progtest{direct-sampling} & $ 0.8029 \pm 0.0001 $ \\
            \progtest{metropolis} & $ 0.8029 \pm 0.0026 $ \\
            \progtest{factor-metropolis} & $ 0.8004 \pm 0.0035 $ \\
            \progtest{factor-metropolis(patch)} & $ 0.8029 \pm 0.0015 $ \\
            \progtest{lifted-metropolis} & $ 0.8033 \pm 0.0003 $ \\
            \progtest{zig-zag} & $ 0.80292 \pm 0.00009 $ \\
            \progtest{factor-zig-zag} & $ 0.8030 \pm 0.0001 $ \\
            \progtest{bounded-lifted} & $ 0.8036 \pm 0.0004 $ \\
            \progtest{bounded-zig-zag} & $ 0.80297 \pm 0.00009 $ \\
            \progtest{bounded-factor-zig-zag} & $ 0.80297 \pm 0.00007 $ \\
            \progtest{bounded-factor-zig-zag(patch)} & $ 0.8029 \pm 0.0001 $ \\
        \end{tabular}
    \end{center}
    \caption{Estimated probability $\prob(x< 0.63)$ for the anharmonic
    oscillator computed by the algorithms discussed in this paper
    (single-$\sigma$ error bars).  }
\label{tab:NumericalResults}
\end{table}
\input{MCMCNutshellArxiv.bbl}

\end{document}

%% file: latexcommands_arxiv.tex
\newcommand{\PROCEDURE}[1]{\textbf{procedure}\ \sub{#1}}

\newcommand{\BRACE}[1]{
\;\;\;  \left\{\begin{array}{l}#1\end{array} \right.}
\newcommand{\IS}[2]{#1 \leftarrow #2}

\newcommand{\FOR}[1]{\textbf{for}\ #1 \textbf{: }}

\newcommand{\WHILE}[1]{\textbf{while } #1 \textbf{: }}
\newcommand{\ENDPROCEDURE}{\text{------} \\ \vspace{-0.8cm}}

\newcommand{\BREAK}{\textbf{break}}
\newcommand{\TRUE}{\textbf{True}}
\newcommand{\IF}[1]{\textbf{if } #1 \textbf{: }}
\newcommand{\ELIF}[1]{\textbf{else if } #1 \textbf{: }}

\newcommand{\AND}{\text{and}}

\newcommand{\ELSE}{\textbf{else: }}

\newcommand{\OUTPUT}[1]{\textbf{output}\ #1}
\newcommand{\PRINT}[1]{\textbf{print}\ #1}

\newcommand{\INPUT}[1]{\textbf{input}\ #1}
\newcommand{\COMMENT}[1]{\text{\footnotesize (#1)}}

\newcommand{\SET}[1]{\{#1\}}

\newcommand{\sub}[1]{\texttt{#1}}
\newcommand{\eq}[1]{eq.~(\ref{#1})}

\newcommand{\Eq}[1]{Eq.~(\ref{#1})}
\newcommand{\Eqq}[1]{Equation~(\ref{#1})}
\newcommand{\eqtwo}[2]{eqs.~(\ref{#1}) and~(\ref{#2})}
\newcommand{\Eqtwo}[2]{Eqs.~(\ref{#1}) and~(\ref{#2})}
\newcommand{\fig}[1]{Fig.~\ref{#1}}
\newcommand{\quot}[1]{``#1''}
\newcommand{\tab}[1]{Table~\ref{#1}} 
 
\newcommand{\app}[1]{App.~\ref{#1}}
\newcommand{\sect}[1]{Sec.~\ref{#1}} 
\newcommand{\subsect}[1]{Sec.~\ref{#1}}

\newcommand{\eg}{\textrm{e.g.}}

\newcommand{\etc}{\textrm{etc.}}

\newcommand{\ie}{\textrm{i.e.}}

\newcommand{\ACAL}{\mathcal{A}}  
\newcommand{\OCAL}{\mathcal{O}}  
\newcommand{\PCAL}{\mathcal{P}}  
\newcommand{\xbar}{\overline{x}}  
\newcommand{\stilde}{\tilde{\sigma}}

\newcommand{\expa}[1]{\mathrm{e}^{#1}}   
\newcommand{\expb}[1]{\exp \glb #1 \grb} 
\newcommand{\expba}[1]{\exp \bigl( #1 \bigr)} 
\newcommand{\expc}[1]{\exp \glc #1 \grc} 
\newcommand{\ran}{\texttt{ran}}
\newcommand{\ranb}[2][]{\ran_{#1} \! \glb #2 \grb}  
\newcommand{\ranc}[2][]{\ran_{#1} \! \glc #2 \grc}  

\newcommand{\loga}[2][]{\log^{#1}\! \gla #2 \gra}  

\newcommand{\minb}[2][]{\min^{#1} \glb #2 \grb}  
\newcommand{\minc}[2][]{\min^{#1} \glc #2 \grc}  

\newcommand{\prob}{\mathbb{P}}

\newcommand{\gla}{\,}  
\newcommand{\gra}{}  
\newcommand{\glb}{\left(}  
\newcommand{\grb}{\right)}  
\newcommand{\glc}{\left[}  
\newcommand{\grc}{\right]}  
\newcommand{\gld}{\left\{}  
\newcommand{\grd}{\right\}}  

\newcommand{\TO}{,\ldots,}

\newcommand{\dd}[1]{\mathrm{d}{#1\ }}   
\newcommand{\ddd}[1]{\mathrm{d}{#1}}   

\newcommand{\qhat}{\hat{q}}
\newcommand{\xhat}{\hat{x}}

\newcommand{\half}{\frac{1}{2}}

\newcommand\diff[1]{\mathrm{d}#1}
\newcommand{\fraca}[2]{#1 / #2}
\newcommand{\fracb}[2]{\frac{#1}{#2}}
\newcommand{\fracc}[2]{\tfrac{#1}{#2}}
\newcommand{\fracd}[2]{\dfrac{#1}{#2}}

%% file: MCMCNutshellArxiv.bbl
%

%% file: MCMCNutshellArxiv.bbl
\begin{thebibliography}{17}%
\makeatletter
\providecommand \@ifxundefined [1]{%
 \@ifx{#1\undefined}
}%
\providecommand \@ifnum [1]{%
 \ifnum #1\expandafter \@firstoftwo
 \else \expandafter \@secondoftwo
 \fi
}%
\providecommand \@ifx [1]{%
 \ifx #1\expandafter \@firstoftwo
 \else \expandafter \@secondoftwo
 \fi
}%
\providecommand \natexlab [1]{#1}%
\providecommand \enquote  [1]{``#1''}%
\providecommand \bibnamefont  [1]{#1}%
\providecommand \bibfnamefont [1]{#1}%
\providecommand \citenamefont [1]{#1}%
\providecommand \href@noop [0]{\@secondoftwo}%
\providecommand \href [0]{\begingroup \@sanitize@url \@href}%
\providecommand \@href[1]{\@@startlink{#1}\@@href}%
\providecommand \@@href[1]{\endgroup#1\@@endlink}%
\providecommand \@sanitize@url [0]{\catcode `\\12\catcode `\$12\catcode
  `\&12\catcode `\#12\catcode `\^12\catcode `\_12\catcode `\%12\relax}%
\providecommand \@@startlink[1]{}%
\providecommand \@@endlink[0]{}%
\providecommand \url  [0]{\begingroup\@sanitize@url \@url }%
\providecommand \@url [1]{\endgroup\@href {#1}{\urlprefix }}%
\providecommand \urlprefix  [0]{URL }%
\providecommand \Eprint [0]{\href }%
\providecommand \doibase [0]{https://doi.org/}%
\providecommand \selectlanguage [0]{\@gobble}%
\providecommand \bibinfo  [0]{\@secondoftwo}%
\providecommand \bibfield  [0]{\@secondoftwo}%
\providecommand \translation [1]{[#1]}%
\providecommand \BibitemOpen [0]{}%
\providecommand \bibitemStop [0]{}%
\providecommand \bibitemNoStop [0]{.\EOS\space}%
\providecommand \EOS [0]{\spacefactor3000\relax}%
\providecommand \BibitemShut  [1]{\csname bibitem#1\endcsname}%
\let\auto@bib@innerbib\@empty
\bibitem [{\citenamefont {Krauth}(2006)}]{SMAC}%
  \BibitemOpen
  \bibfield  {author} {\bibinfo {author} {\bibfnamefont {W.}~\bibnamefont
  {Krauth}},\ }\href@noop {} {\emph {\bibinfo {title} {{Statistical Mechanics:
  Algorithms and Computations}}}}\ (\bibinfo  {publisher} {Oxford University
  Press},\ \bibinfo {year} {2006})\BibitemShut {NoStop}%
\bibitem [{\citenamefont {Tartero}\ \emph {et~al.}()\citenamefont {Tartero},
  \citenamefont {Vionnet},\ and\ \citenamefont {Krauth}}]{Tartero2023b}%
  \BibitemOpen
  \bibfield  {author} {\bibinfo {author} {\bibfnamefont {G.}~\bibnamefont
  {Tartero}}, \bibinfo {author} {\bibfnamefont {S.}~\bibnamefont {Vionnet}},\
  and\ \bibinfo {author} {\bibfnamefont {W.}~\bibnamefont {Krauth}},\
  }\bibfield  {title} {\bibinfo {title} {{Fast sampling of Lennard-Jones
  systems without cutoffs}},\ }\bibinfo {note} {manuscript in
  preparation}\BibitemShut {NoStop}%
\bibitem [{\citenamefont {Davis}(1984)}]{Davis1984}%
  \BibitemOpen
  \bibfield  {author} {\bibinfo {author} {\bibfnamefont {M.~H.~A.}\
  \bibnamefont {Davis}},\ }\bibfield  {title} {\bibinfo {title}
  {{Piecewise-Deterministic Markov Processes: A General Class of Non-Diffusion
  Stochastic Models}},\ }\href
  {https://doi.org/https://doi.org/10.1111/j.2517-6161.1984.tb01308.x}
  {\bibfield  {journal} {\bibinfo  {journal} {J. R. Stat. Soc. Series B Stat.
  Methodol.}\ }\textbf {\bibinfo {volume} {46}},\ \bibinfo {pages} {353}
  (\bibinfo {year} {1984})}\BibitemShut {NoStop}%
\bibitem [{\citenamefont {Landau}\ and\ \citenamefont
  {Lifshitz}(1982)}]{LandauMechanics1982}%
  \BibitemOpen
  \bibfield  {author} {\bibinfo {author} {\bibfnamefont {L.}~\bibnamefont
  {Landau}}\ and\ \bibinfo {author} {\bibfnamefont {E.}~\bibnamefont
  {Lifshitz}},\ }\href@noop {} {\emph {\bibinfo {title} {Mechanics: Volume
  1}}},\ \bibinfo {number} {vol.~1}\ (\bibinfo  {publisher} {Elsevier
  Science},\ \bibinfo {year} {1982})\BibitemShut {NoStop}%
\bibitem [{\citenamefont {Li}\ \emph {et~al.}(2022)\citenamefont {Li},
  \citenamefont {Nishikawa}, \citenamefont {H\"{o}llmer}, \citenamefont
  {Carillo}, \citenamefont {Maggs},\ and\ \citenamefont {Krauth}}]{Li2022}%
  \BibitemOpen
  \bibfield  {author} {\bibinfo {author} {\bibfnamefont {B.}~\bibnamefont
  {Li}}, \bibinfo {author} {\bibfnamefont {Y.}~\bibnamefont {Nishikawa}},
  \bibinfo {author} {\bibfnamefont {P.}~\bibnamefont {H\"{o}llmer}}, \bibinfo
  {author} {\bibfnamefont {L.}~\bibnamefont {Carillo}}, \bibinfo {author}
  {\bibfnamefont {A.~C.}\ \bibnamefont {Maggs}},\ and\ \bibinfo {author}
  {\bibfnamefont {W.}~\bibnamefont {Krauth}},\ }\bibfield  {title} {\bibinfo
  {title} {{Hard-disk pressure computations{\textemdash}a historic
  perspective}},\ }\href {https://doi.org/10.1063/5.0126437} {\bibfield
  {journal} {\bibinfo  {journal} {The Journal of Chemical Physics}\ }\textbf
  {\bibinfo {volume} {157}},\ \bibinfo {pages} {234111} (\bibinfo {year}
  {2022})}\BibitemShut {NoStop}%
\bibitem [{\citenamefont {{Michel}}\ \emph {et~al.}(2014)\citenamefont
  {{Michel}}, \citenamefont {{Kapfer}},\ and\ \citenamefont
  {{Krauth}}}]{Michel2014JCP}%
  \BibitemOpen
  \bibfield  {author} {\bibinfo {author} {\bibfnamefont {M.}~\bibnamefont
  {{Michel}}}, \bibinfo {author} {\bibfnamefont {S.~C.}\ \bibnamefont
  {{Kapfer}}},\ and\ \bibinfo {author} {\bibfnamefont {W.}~\bibnamefont
  {{Krauth}}},\ }\bibfield  {title} {\bibinfo {title} {{Generalized event-chain
  Monte Carlo: Constructing rejection-free global-balance algorithms from
  infinitesimal steps}},\ }\href {https://doi.org/10.1063/1.4863991} {\bibfield
   {journal} {\bibinfo  {journal} {J. Chem. Phys.}\ }\textbf {\bibinfo {volume}
  {140}},\ \bibinfo {eid} {054116} (\bibinfo {year} {2014})}\BibitemShut
  {NoStop}%
\bibitem [{\citenamefont {Diaconis}\ \emph {et~al.}(2000)\citenamefont
  {Diaconis}, \citenamefont {Holmes},\ and\ \citenamefont
  {Neal}}]{Diaconis2000}%
  \BibitemOpen
  \bibfield  {author} {\bibinfo {author} {\bibfnamefont {P.}~\bibnamefont
  {Diaconis}}, \bibinfo {author} {\bibfnamefont {S.}~\bibnamefont {Holmes}},\
  and\ \bibinfo {author} {\bibfnamefont {R.~M.}\ \bibnamefont {Neal}},\
  }\bibfield  {title} {\bibinfo {title} {{Analysis of a nonreversible Markov
  chain sampler}},\ }\href {https://doi.org/10.1214/aoap/1019487508} {\bibfield
   {journal} {\bibinfo  {journal} {Ann. Appl. Probab.}\ }\textbf {\bibinfo
  {volume} {10}},\ \bibinfo {pages} {726} (\bibinfo {year} {2000})}\BibitemShut
  {NoStop}%
\bibitem [{\citenamefont {Chen}\ \emph {et~al.}(1999)\citenamefont {Chen},
  \citenamefont {Lovász},\ and\ \citenamefont {Pak}}]{Chen1999}%
  \BibitemOpen
  \bibfield  {author} {\bibinfo {author} {\bibfnamefont {F.}~\bibnamefont
  {Chen}}, \bibinfo {author} {\bibfnamefont {L.}~\bibnamefont {Lovász}},\ and\
  \bibinfo {author} {\bibfnamefont {I.}~\bibnamefont {Pak}},\ }\bibfield
  {title} {\bibinfo {title} {{Lifting Markov Chains to Speed up Mixing}},\
  }\href@noop {} {\bibfield  {journal} {\bibinfo  {journal} {Proceedings of the
  17th Annual ACM Symposium on Theory of Computing}\ ,\ \bibinfo {pages} {275}}
  (\bibinfo {year} {1999})}\BibitemShut {NoStop}%
\bibitem [{\citenamefont {Bierkens}\ \emph {et~al.}(2019)\citenamefont
  {Bierkens}, \citenamefont {Fearnhead},\ and\ \citenamefont
  {Roberts}}]{Bierkens2019}%
  \BibitemOpen
  \bibfield  {author} {\bibinfo {author} {\bibfnamefont {J.}~\bibnamefont
  {Bierkens}}, \bibinfo {author} {\bibfnamefont {P.}~\bibnamefont
  {Fearnhead}},\ and\ \bibinfo {author} {\bibfnamefont {G.}~\bibnamefont
  {Roberts}},\ }\bibfield  {title} {\bibinfo {title} {{The Zig-Zag process and
  super-efficient sampling for Bayesian analysis of big data}},\ }\href
  {https://doi.org/10.1214/18-AOS1715} {\bibfield  {journal} {\bibinfo
  {journal} {Ann. Stat.}\ }\textbf {\bibinfo {volume} {47}},\ \bibinfo {pages}
  {1288} (\bibinfo {year} {2019})}\BibitemShut {NoStop}%
\bibitem [{\citenamefont {Levin}\ \emph {et~al.}(2008)\citenamefont {Levin},
  \citenamefont {Peres},\ and\ \citenamefont {Wilmer}}]{Levin2008}%
  \BibitemOpen
  \bibfield  {author} {\bibinfo {author} {\bibfnamefont {D.~A.}\ \bibnamefont
  {Levin}}, \bibinfo {author} {\bibfnamefont {Y.}~\bibnamefont {Peres}},\ and\
  \bibinfo {author} {\bibfnamefont {E.~L.}\ \bibnamefont {Wilmer}},\
  }\href@noop {} {\emph {\bibinfo {title} {{Markov Chains and Mixing Times}}}}\
  (\bibinfo  {publisher} {American Mathematical Society},\ \bibinfo {year}
  {2008})\BibitemShut {NoStop}%
\bibitem [{\citenamefont {Krauth}(2021)}]{Krauth2021eventchain}%
  \BibitemOpen
  \bibfield  {author} {\bibinfo {author} {\bibfnamefont {W.}~\bibnamefont
  {Krauth}},\ }\bibfield  {title} {\bibinfo {title} {{Event-Chain Monte Carlo:
  Foundations, Applications, and Prospects}},\ }\href
  {https://doi.org/10.3389/fphy.2021.663457} {\bibfield  {journal} {\bibinfo
  {journal} {Front. Phys.}\ }\textbf {\bibinfo {volume} {9}},\ \bibinfo {pages}
  {229} (\bibinfo {year} {2021})}\BibitemShut {NoStop}%
\bibitem [{\citenamefont {Peters}\ and\ \citenamefont
  {de~With}(2012)}]{Peters_2012}%
  \BibitemOpen
  \bibfield  {author} {\bibinfo {author} {\bibfnamefont {E.~A. J.~F.}\
  \bibnamefont {Peters}}\ and\ \bibinfo {author} {\bibfnamefont
  {G.}~\bibnamefont {de~With}},\ }\bibfield  {title} {\bibinfo {title}
  {{Rejection-free Monte Carlo sampling for general potentials}},\ }\href
  {https://doi.org/10.1103/PhysRevE.85.026703} {\bibfield  {journal} {\bibinfo
  {journal} {Phys. Rev. E}\ }\textbf {\bibinfo {volume} {85}},\ \bibinfo
  {pages} {026703} (\bibinfo {year} {2012})}\BibitemShut {NoStop}%
\bibitem [{\citenamefont {Lewis}\ and\ \citenamefont
  {Shedler}(1979)}]{LewisShedler1979}%
  \BibitemOpen
  \bibfield  {author} {\bibinfo {author} {\bibfnamefont {P.~A.~W.}\
  \bibnamefont {Lewis}}\ and\ \bibinfo {author} {\bibfnamefont {G.~S.}\
  \bibnamefont {Shedler}},\ }\bibfield  {title} {\bibinfo {title} {{Simulation
  of nonhomogeneous Poisson processes by thinning}},\ }\href
  {https://doi.org/10.1002/nav.3800260304} {\bibfield  {journal} {\bibinfo
  {journal} {Naval Research Logistics Quarterly}\ }\textbf {\bibinfo {volume}
  {26}},\ \bibinfo {pages} {403} (\bibinfo {year} {1979})}\BibitemShut
  {NoStop}%
\bibitem [{\citenamefont {Kapfer}\ and\ \citenamefont
  {Krauth}(2016)}]{KapferKrauth2016}%
  \BibitemOpen
  \bibfield  {author} {\bibinfo {author} {\bibfnamefont {S.~C.}\ \bibnamefont
  {Kapfer}}\ and\ \bibinfo {author} {\bibfnamefont {W.}~\bibnamefont
  {Krauth}},\ }\bibfield  {title} {\bibinfo {title} {{Cell-veto Monte Carlo
  algorithm for long-range systems}},\ }\href
  {https://doi.org/10.1103/PhysRevE.94.031302} {\bibfield  {journal} {\bibinfo
  {journal} {Phys. Rev. E}\ }\textbf {\bibinfo {volume} {94}},\ \bibinfo
  {pages} {031302} (\bibinfo {year} {2016})}\BibitemShut {NoStop}%
\bibitem [{\citenamefont {Walker}(1977)}]{Walker1977AnEfficientMethod}%
  \BibitemOpen
  \bibfield  {author} {\bibinfo {author} {\bibfnamefont {A.~J.}\ \bibnamefont
  {Walker}},\ }\bibfield  {title} {\bibinfo {title} {{An Efficient Method for
  Generating Discrete Random Variables with General Distributions}},\ }\href
  {https://doi.org/10.1145/355744.355749} {\bibfield  {journal} {\bibinfo
  {journal} {ACM Trans. Math. Softw.}\ }\textbf {\bibinfo {volume} {3}},\
  \bibinfo {pages} {253} (\bibinfo {year} {1977})}\BibitemShut {NoStop}%
\bibitem [{NIS()}]{NIST}%
  \BibitemOpen
  \href@noop {} {\bibinfo {title} {{NIST} {D}igital {L}ibrary of {M}athematical
  {F}unctions}},\ \bibinfo {howpublished} {\url{http://dlmf.nist.gov/}},\
  \bibinfo {note} {edited by {F}. {W}. {J}. {O}lver, {A}. {B}. {O}lde
  {D}aalhuis, {D}. {W}. {L}ozier, {B}. {I}. {S}chneider, {R}. {F}.
  {B}oisvert, {C}. {W}. {C}lark, {B}. {R}. {M}iller, {B}. {V}. {S}aunders,
  {H}. {S}. {C}ohl, and {M}. {A}. {M}c{C}lain}\BibitemShut {NoStop}%
\bibitem [{Note1()}]{Note1}%
  \BibitemOpen
  \bibinfo {note} {The url of the repository is \protect \url
  {https://github.com/jellyfysh/MCMCNutshell}.}\BibitemShut {Stop}%
\end{thebibliography}
